\documentclass[aps,showpacs,preprint,epsf,epsfig, nofootinbib]{revtex4-1}
\usepackage{graphicx}
\usepackage{epsfig,graphicx,color,appendix}
\usepackage{multirow}
\usepackage{tabularx}
\usepackage{capt-of}
\usepackage{amsmath}
\usepackage{caption}
\usepackage[countmax]{subfloat}

\begin{document}
\def\ben{\begin{eqnarray}}
\def\en{\end{eqnarray}}
\def\non{\nonumber}
\def\la{\langle}
\def\ra{\rangle}
\def\t{\times}
\def\pp{{\prime\prime}}
\def\nc{N_c^{\rm eff}}
\def\vp{\varepsilon}
\def\hep{\hat{\varepsilon}}
\def\a{{\cal A}}
\def\B{{\cal B}}
\def\c{{\cal C}}
\def\d{{\cal D}}
\def\e{{\cal E}}
\def\p{{\cal P}}
\def\tt{{\cal T}}
\def\up{\uparrow}
\def\dw{\downarrow}
\def\vma{{_{V-A}}}
\def\vpa{{_{V+A}}}
\def\smp{{_{S-P}}}
\def\spp{{_{S+P}}}
\def\J{{J/\psi}}
\def\ov{\overline}
\def\Lqcd{{\Lambda_{\rm QCD}}}
\def\pr{{Phys. Rev.}~}
\def\prl{{ Phys. Rev. Lett.}~}
\def\pl{{ Phys. Lett.}~}
\def\np{{ Nucl. Phys.}~}
\def\zp{{ Z. Phys.}~}
\long\def\symbolfootnote[#1]#2{\begingroup%
\def\thefootnote{\fnsymbol{footnote}}\footnote[#1]{#2}\endgroup}
\def\lsim{ {\ \lower-1.2pt\vbox{\hbox{\rlap{$<$}\lower5pt\vbox{\hbox{$\sim$}
}}}\ } }
\def\gsim{ {\ \lower-1.2pt\vbox{\hbox{\rlap{$>$}\lower5pt\vbox{\hbox{$\sim$}
}}}\ } }

\font\el=cmbx10 scaled \magstep2{\obeylines\hfill \today}

\vskip 1.5 cm

\centerline{\large\bf Branching ratios of $\pmb B_c$  Meson Decaying }
 \centerline{\large\bf to Vector and Axial-Vector Mesons}
\small
\vskip 1.0 cm

\centerline{\bf Rohit Dhir \symbolfootnote[2]{dhir.rohit@gmail.com} and C. S. Kim \symbolfootnote[3]{cskim@yonsei.ac.kr,~~~ Corresponding Author} }
\medskip
\centerline{\it Department of Physics and IPAP, Yonsei University, Seoul 120-749, Korea}
\bigskip
\bigskip
\begin{center}
{\large \bf Abstract}
\end{center}
We investigate the weak decays of $B_c$ mesons in Cabibbo-Kobayashi-Maskawa favored and suppressed modes.
We present a detailed analysis of the $B_c$ meson decaying to vector meson ($V$) and axial-vector meson ($A$) in the final state.
We also give the form factors involving $B_{c} \to A$ transition in the Isgur-Scora-Grinstein-Wise II framework and consequently,
predict the branching ratios of $B_{c} \to VA$ and $AA$ decays.

\medskip
\medskip

\vfill

\section{Introduction}

 The $B_c$  meson was first discovered by CDF collaboration  at Fermilab \cite{1} in 1998. At present, a more precise measurement of its mass and life time is available in Particle Data Group (PDG) \cite{2} \textit{i.e.} $M_{B_{c}}=6.277 \pm 0.006 $ GeV and $\tau_{B_{c}}= (0.453 \pm 0.042) \t 10^{-12}$ s. It is believed that LHC-b is expected to produce $5 \t 10^{10}$ events per year \cite{3,4,5,6}, which is around $10 \%$ of the total $B$ meson data. This will provide a rich amount of information regarding $B_c$ meson.

 The $B_c$ meson is a unique Standard Model (SM) particle which is quark-antiquark bound state ($b \bar{c}$) consisting two heavy quarks of different flavors and, therefore, is flavor asymmetric. The study of $B_c$ meson is of special interest as compared to the flavor-neutral heavy quarkonium ($b \bar{b}$, $c\bar{c}$) states, as it only decays \textit{via} weak interactions, while the later predominantly decays via strong interactions and/or electromagnetic interactions.
 The decay processes of the $B_c$ meson can be divided into three categories involving: (\textit{i}) decay of the \textit{b} quark with \textit{c}-quark being spectator, (\textit{ii})  decay  of the \textit{c} quark with \textit{b}-quark being spectator, (\textit{iii}) the relatively suppressed annihilation of \textit{b} and $\bar{c}$ which is ignored in present work.
One can find several theoretical works based on a variety of quark models  \cite{7, 8, 9, 10, 11, 12, 13, 14, 15,16,17,18} for the semileptonic and nonleptonic decays of $B_c$ emitting \textit{s}-wave mesons, pseudoscalar (\textit{P}) and vector (\textit{V}) mesons. A relatively less attention has been paid to the \textit{p}-wave meson emitting weak decays of $B_c$ meson \cite{19,20,21,22,23,24,25}. In recent past, several relativistic and non relativistic quark models \cite{13,15,19,20,21,22} are used employing factorization approach to calculate branching ratios (BRs) of $B_c$ meson decaying to a \textit{p}-wave charmonium ($ c \bar{c} $) in the final state. Most recently, Salpeter Method \cite{24} and Improved Bethe-Salpeter Approach \cite{25} are used to probe non-leptonic decays of $B_c$ meson. On experimental side,  more measurements regarding $B_c$ meson will be available soon at Large Hadron Collider (LHC), LHC-b and Super-B experiments. A high precision instrumentation at these experiments may provide precise measurement of BRs of the order of ($10^{-6}$), which makes study of $B_c$ meson decays more interesting.
The developing theoretical and experimental aspects of the $B_c$ meson physics motivate us to investigate weak hadronic decays of $B_c$ meson emitting vector (\textit{V}) and axial-vector (\textit{A}) mesons in the final state. We employ the improved Isgur-Scora-Grinstein-Wise quark model (known as ISGW II Model) \cite{26,27} to obtain $B_{c} \to A$ transition form factors. Using the factorization approach, we calculate the decay amplitudes and predict branching ratios of $B_c \to VA/AA$ decays. For $B_c \to V$ transition form factors we rely on our previous work \cite{18} based on flavor dependence effects in
Bauer-Stech-Wirbel (BSW) model frame work \cite{28}.

The presentation of the article goes as follows. We discuss the mass spectrum and the methodology in Sections II and III, respectively. Decay constants are discussed in Section IV. We present the $B_{c} \to A$ transitions form factor in ISGW II Model and give a brief account for $B_c \to V$ transitions form factors in Section V, respectively. Consequently, the branching ratios are estimated. Results and discussions are presented in Section VI and last Section contains summary and conclusions.

\section{Mass Spectrum}
Two types of axial-vector mesons exist, ${}^{3} P_{1} $($J^{PC} =1^{++} $) and ${}^{1} P_{1} $ $(J^{PC} =1^{+-} )$, with respect to the quark model $q\bar{q}$ assignments. These states can exhibit two kinds of mixing behavior: one, mixing between ${}^{3} P_{1} $ or ${}^{1} P_{1} $ states themselves; second, mixing among ${}^{3} P_{1} $ or ${}^{1} P_{1} $ states.  The following non-strange and uncharmed mesons states have been observed experimentally \cite{2}:

\begin{enumerate}
\item[a)]   ${}^{3} P_{1} $ multiplet consists: isovector $a_{1}(1.230) $  and four isoscalars $f_{1} (1.285)$, $f_{1} (1.420)$, $f'_{1} (1.512) $ and $\chi _{c1} (3.511)$;

 \item[b)] ${}^{1} P_{1} $ multiplet consists: isovector $b_{1} (1.229)$ and three isoscalars $ h_{1} (1.170)$, $ h'_{1} (1.380)$ and $h_{c1} (3.526)$, where spin and parity of the $h_{c1} (3.526)$ and C-parity of $ h'_{1} (1.380)$ remain to be confirmed\footnote{ Here quantities in the brackets indicate their respective masses (in GeV).}.
\end{enumerate}

In the present work, we use the following mixing scheme for the isoscalar ($1^{++} $) mesons:

\ben
 f_{1} (1.285) &=& \frac{1}{\sqrt{2} } (u\overline{u}+d\overline{d})\cos \phi_{A} +(s\overline{s})\sin \phi_{A} \non
  \\ f'_{1} (1.512)\,&=&\frac{1}{\sqrt{2} } (u\overline{u}+d\overline{d})\sin \phi_{A} -(s\overline{s})\cos \phi _{A} \non
 \\ \chi _{c1} (3.511)&=&(c\bar{c}).
\en
Likewise, mixing for isoscalar ($1^{+-} $) mesons is given by

\ben
 h_{1} (1.170)&=&\frac{1}{\sqrt{2} } (u\overline{u}+d\overline{d})\cos \phi {}_{A'} +(s\overline{s})\sin \phi {}_{A'}, \non \\  h'_{1} (1.380)&=&\frac{1}{\sqrt{2} } (u\overline{u}+d\overline{d})\sin \phi _{A'} -(s\overline{s})\cos \phi _{A'} , \non
\\h_{c1} (3.526)&=&(c\bar{c}),
\en
with
\[\phi _{A(A')} =  \theta (ideal) - \theta _{A(A')} (physical).\]

It has been observed that $f_{1}(1.285) \to 4\pi $/$\eta \pi \pi $, $f_{1}^{'} (1.512) \to K\bar{K}\pi $, $h_{1} (1.170) \to \rho \pi $ and $h_{1}^{'} \to K\bar{K}^{*} $/$\bar{K}K^{*} $ predominantly, which seems to favor the ideal mixing for both $1^{++} $ and $1^{+-} $ nonets $i.e.$,

\begin{equation}
\phi _{A}  =\phi _{A'}  = 0^{\circ }.
\end{equation}

The  hidden-flavor diagonal ${}^{3} P_{1} $ and ${}^{1} P_{1} $   states have opposite C-parity and therefore, cannot mix. However, their is no restriction on such mixing in strange and charmed states, which are most likely a mixture of ${}^{3} P_{1} $ and ${}^{1} P_{1} $ states.
States involving strange partners of $A\, (J^{PC} =1^{++} )$ and $A'(J^{PC} =1^{+-} )$ states $i.e.$ $K_{1A} $ and $K_{1A'} $ mesons mix to generate the physical states in the following manner:

\ben
 K_{1} (1.270)&=& K_{1A}  \sin \theta _{K}  + K_{1A'}  \cos \theta _{K} ,\non
  \\ \underline{K}_{1} (1.400)&=& K_{1A}  \cos \theta _{K}  - K_{1A'}  \sin \theta _{K} .
\en

  Numerous analysis based on phenomenological studies indicate that strange axial vector meson states mixing angle $\theta_{K}$ lies in the vicinity of $\sim 35^{\circ} $ and $\sim 55^{\circ}$, see for details \cite{29}. Experimental information based on $\tau \to K_{1} (1.270)$ / $K_{1} (1.400)+ \nu _{\tau } $ data yields $\theta _{K} =\pm \, 37^{\circ} $ and $\theta _{K} =\pm \, 58^{\circ} $ \cite{30}. However, the negative mixing angle solutions are favored by $D\to K_{1} (1.270) \pi$ /$ K_{1} (1.400) \pi $ decays and experimental measurement of the ratio of $K_1 \gamma$ production in $B$ decays \cite{31}. Following the discussions given in Ref. \cite{29}, which states that mixing angle $\theta_{K} \sim 35^{\circ}$ is preferred over $\sim 55^{\circ}$, we use $\theta_{K} = -37 ^{\circ}$ in our numerical calculations. It is based on the observation that choice of angle for $f-f^{'}$ and $h-h^{'}$ mixing schemes (which are close to ideal mixing) are intimately related to choice of mixing angle $\theta_{K}$.

In general, mixing of charmed and strange charmed states is given by

\ben
 D_{1} (2.427)&=& D_{1A}  \sin \theta _{D_{1} }  + D_{1A'}  \cos \theta _{D_{1} } , \non \\ \underline{D}_{1} (2.422)&=& D_{1A}  \cos \theta _{D_{1} }  - D_{1A'}  \sin \theta _{D_{1} } ,
\en
and
\ben
 D_{s1} (2.460)&=& D_{s1A}  \sin \theta _{D_{s1} } + D_{s1A'}  \cos \theta _{D_{s1} } ,\non \\ \underline{D}_{s1} (2.535)&=& D_{s1A}  \cos \theta _{D_{s1} }  - D_{s1A'}  \sin \theta _{D_{s1} } ,
\en

As pointed out in \cite{31}, for heavy mesons the heavy quark spin $S_{Q} $  and the total angular momentum of the light antiquark can be used as good quantum numbers, separately. In the heavy quark limit, the physical mass eigenstates $P_{1}^{3/ 2}  $ and $P_{1}^{1/2} $ with $J^{P} =1^{+} $ can be expressed as a combination of ${}^{3} P_{1} $ and ${}^{1} P_{1} $ states as

\ben
 |P_{1}^{1/2 } > &=& -\sqrt{\frac{1}{3} } |{}^{1} P_{1} > + \sqrt{\frac{2}{3} } |{}^{3} P_{1} >,\non \\ |P_{1}^{3/2} >&=& \sqrt{\frac{2}{3} } |{}^{1} P_{1} >+\sqrt{\frac{1}{3} } |{}^{3} P_{1} >.
 \en
 Thus, the states $D_{1} (2.427)$ and $\underline{D}_{1} (2.422)$ can be identified as $P_{1}^{1/2} $ and $P_{1}^{3/2} $, respectively. However, beyond the heavy quark limit, there is a mixing between $P_{1}^{1/2} $ and $P_{1}^{3/2} $ given by

\ben
D_{1} (2.427)&=& D_{1}^{1/2} \cos \theta _{2}  + D_{1}^{3/2}  \sin \theta _{2} , \non
\\
\underline{D}_{1} (2.422)&=& -D_{1}^{1/2}  \sin \theta _{2}  + D_{1}^{3/2}  \cos \theta _{2} .
\en
Similarly, for strange charmed axial-vector mesons,

\ben D_{s1} (2.460)&=& D_{s1}^{1/2} \cos \theta _{3}  + D_{s1}^{3/2}  \sin \theta _{3} ,\non \\ \underline{D}_{s1} (2.535)&=& -D_{s1}^{1/2 }  \sin \theta _{3}  +D_{s1}^{3/2 } \cos \theta _{3} .
\en
A detailed analysis by Belle \cite{32} yields the mixing angle $\theta _{2} =(-5.7\pm 2.4)^{\circ } $.  while the quark potential model \cite{33, 34} determines $\theta _{3} \approx 7^{\circ } $.

For $\omega $ and $\phi$ vector mesons states, we consider ideal mixing $i.e.$
$\omega =  \frac{1}{\sqrt{2} } (u\overline{u}+d\overline{d})$ and $
 \phi = \frac{1}{\sqrt{2} } (s\overline{s})$ \cite{2}.

\section{Methodology}
\subsection{Weak Hamiltonian}

 The QCD modified weak Hamiltonian \cite{35} generating the $B_c$ decay involving $b\to c$ transition in
 Cabibbo-Kobayashi-Maskawa (CKM) enhanced modes ($\Delta b = 1, \Delta C =1, \Delta S=0 ;~~\Delta b = 1, \Delta C =0,\Delta S =-1$) is given by

\ben
 H_{w}^{\Delta b=1} =\frac{G_{F} }{\sqrt{2} } \{ V_{cb} V_{ud}^{*} [c_{1} (\mu )(\bar{c}b)(\bar{d}u)+c_{2} (\mu )(\bar{c}u)(\bar{d}b)]\non
 \\ + V_{cb} V_{cs}^{*} [c_{1} (\mu )(\bar{c}b)(\bar{s}c)+c_{2} (\mu )(\bar{c}c)(\bar{s}b)] \non\\ +V_{cb} V_{us}^{*} [c_{1} (\mu )(\bar{c}b)(\bar{s}u)+c_{2} (\mu )(\bar{c}u)(\bar{s}b)]\non \\+V_{cb} V_{cd}^{*} [c_{1} (\mu )(\bar{c}b)(\bar{d}c)+c_{2} (\mu )(\bar{c}c)(\bar{d}b)] \}
 \en
and CKM suppressed ($\Delta b = 1, \Delta C=1, \Delta S=-1; ~~\Delta b = 1, \Delta C=1, \Delta S=1; ~~\Delta b = 1, \Delta C=-1, \Delta S=-1; ~~\Delta b = 1, \Delta C=-1, \Delta S=0$) $b \to u$ transitions is given by

\ben
H_{w}^{\Delta b=1} =\frac{G_{F} }{\sqrt{2} } \{ V_{ub} V_{cs}^{*} [c_{1} (\mu )(\bar{u}b)(\bar{s}c)+c_{2} (\mu )(\bar{s}b)(\bar{u}c)]\non
 \\ + V_{ub} V_{ud}^{*} [c_{1} (\mu )(\bar{u}b)(\bar{d}u)+c_{2} (\mu )(\bar{d}b)(\bar{u}u)] \non\\ +V_{ub} V_{us}^{*} [c_{1} (\mu )(\bar{u}b)(\bar{s}u)+c_{2} (\mu )(\bar{s}b)(\bar{u}u)]\non \\+V_{ub} V_{cd}^{*} [c_{1} (\mu )(\bar{u}b)(\bar{d}c)+c_{2} (\mu )(\bar{u}c)(\bar{d}b)] \}
\en
 where $\bar{q}q \equiv \bar{q} \gamma_{\mu}(1-\gamma_5)q$, $G_{F} $ is the Fermi constant and $V_{ij} $ are the CKM matrix elements, $c_{1} $ and $c_{2} $ are the standard perturbative QCD coefficients, usually taken at $\mu \approx m_{b}^{2} $.
  In addition to the bottom changing decays, the bottom
conserving decay channel is also available for the $B_c$ meson, where the charm quark decays to an $s$ or a $d$ quark. However, in case of $B_c \to VA/AA$ decays, these modes are kinematically forbidden.

\subsection {Decay Amplitudes}

\par In generalized factorization hypothesis the decay amplitudes can be expressed as a product of the matrix elements of weak currents (up to the weak scale factor of $\frac{G_{F} }{\sqrt{2}} $ $\times$ CKM elements $\times$ QCD factor) given by

\ben
 \la PA  | H_{w}  | B_{c} \ra  \sim  \la  P  | J^{\mu}  | 0 \ra  \la  A  | J_{\mu }  | B_{c} \ra  & + & \la  A  | J^{\mu } | 0 \ra  \la  P | J_{\mu }  | B_{c}  \ra , \non \\
\la  PA' | H_{w}   | B_{c} \ra  \sim  \la  P  | J^{\mu }  | 0 \ra \la A' | J_{\mu } | B_{c}  \ra  &+& \la A'  | J^{\mu }  | 0 \ra  \la  P  | J_{\mu }  | B_{c}  \ra.
 \en
Using Lorentz invariance, the hadronic transition matrix elements \cite{26, 27, 28} for the relevant weak current between meson states can be expressed as

\ben
\la V(k_V)| A_{\mu} |0 \ra &=& \vp_{\mu}^{*} f_{V} m_{V}, \non\\
\la{V}(k_V, \vp)|V_\mu |{B_{c}} (k_{B_{c}})\ra
 &=& -i \frac{2}{m_{B_{c}} + m_{V}} \epsilon_{\mu\nu\alpha\beta}
 \vp^{*\nu}
 k_{B_{c}}^\alpha k_V^{\beta} V^{B_{c}V}(q^2),
\nonumber \\
 \la V (k_V,\vp)|A_\mu|{B_{c}}(k_{B_{c}})\ra
 &=& (m_{B_{c}} + m_{V}) \vp^{*}_{\mu} A_1^{B_{c}V}(q^2)
 - (\vp^{*} \cdot k_{B_{c}})
(k_{B_{c}} + k)_\mu \frac{A_2^{B_{c}V}(q^2)}{m_{B_{c}} + m_{V}}
\nonumber \\
&& - 2 m_{V} \frac{\vp^{*}\cdot k_{B_{c}}} {q^2} {q^\mu}
\left[A_3^{B_{c}V}(q^2) - A_0^{B_{c}V}(q^2)\right],
\en
where $q = (k_{B_{c}}- k_{V})_\mu$, $V_3^{B_{c}V}(0) = V_0^{B_{c}V}(0)$ and

\ben
A_3^{B_{c}V}(q^2) &=&  \frac{m_{B_{c}} + m_{V}}{2 m_{V}} A_1^{B_{c}V}(q^2) - \frac{m_{B_{c}}
 - m_{V}}{2 m_{V}} A_2^{B_{c} V}(q^2).
 \en
 Similarly, for axial vector meson states:
\ben
\la A(k_{A} ,\vp ) | A_{\mu } | 0 \ra  &=& \vp _{\mu }^{*} m_{A} f_{A} ,\non  \\
\la A'(k_{A'} ,\vp ) \left|  A_{\mu } \right| 0 \ra  &=& \vp _{\mu }^{*} m_{A'} f_{A'} , \\
\la A(k_{A} ,\vp ) \left| V_{\mu } \right| B_{c} (k_{B_{c} } ) \ra &=& l\vp _{\mu }^{*} + c_{+} (\vp ^{*} \cdot k_{B_{c} } )(k_{B_{c} } + k_{A} )_{\mu } + c_{-} (\vp ^{*} \cdot k_{B_{c} } )(k_{B_{c} } -k_{A} )_{\mu } , \non \\
\la{A}(k_A, \varepsilon)|A_\mu|{B_c} (k_{B_{c}})\ra &=& i q' \epsilon_{\mu\nu\alpha\beta} \varepsilon^{*\nu}(k_{B_{c}}+k_A)^\alpha (k_{B_{c}}-k_A)^{\beta} ,\non \\
\la A'(k_{A'} ,\vp ) \left| V_{\mu } \right| B_{c} (k_{B_{c} } ) \ra  &=& r\vp _{\mu }^{*} +s_{+} (\vp ^{*} \cdot k_{B_{c} } )(k_{B_{c} } +k_{A'} )_{\mu } +s_{-} (\vp ^{*} \cdot k_{B_{c} } )(k_{B_{c} } -k_{A'} )_{\mu } ,\non \\
\la{A'}(k_A, \varepsilon)|A_\mu|{B_c} (k_{B_{c}})\ra &=& i v \epsilon_{\mu\nu\alpha\beta}\varepsilon^{*\nu}(k_{B_{c}}+k_{A'})^\alpha (k_{B_{c}}-k_{A'})^{\beta},
\en
where $q_{\mu } =(k_{B_{c} } -k_{A} )_{\mu }$.

It may be noted that $B_c \to A/A'$ transition form factors in ISGW II framework are related to BSW type form factor \cite{28} notations $i.e.$ $A$, $V_{0,1,2} $ as follows
\ben
 A(q^2)&=& -q'(q^2)(m_{B_{c}}+m_A);\non \\
V_1(q^2)&=& l (q^2)/(m_{B_{c}}+m_A);\non \\
V_2(q^2)&=& -c_+(q^2)(m_{B_{c}}+m_A);\non \\
V_0(q^2) &=& \frac{1}{(2m_A)}[(m_{B_{c}} + m_A)V_1(q^2) - (m_{B_{c}} - m_A)V_2(q^2)-q^2 c_-(q^2)].
 \en

Sandwiching the weak Hamiltonian (3.1) and (3.2) between the initial and the final states, the decay amplitudes for various $B_{c} \to MA$ decay modes ($M$ = $V$ or $A$) can be obtained for the following three categories \cite{28}:

\begin{enumerate}
\item  Class I transitions: contain those decays which are caused by color favored diagram and the decay amplitudes are proportional to $a_{1} $, where $a_{1} (\mu )=c_{1} (\mu )+\frac{1}{N_{c} } \, c_{2} (\mu )$, and $N_{c} $ is the number of colors.

\item  Class II transitions: consist of those decays which are caused by color suppressed diagrams. The decay amplitude in this class is proportional to $a_{2} $ \textit{i.e.} for the color suppressed modes  $a_{2} (\mu )=c_{2} (\mu )+\frac{1}{N_{c} } \, c_{1} (\mu ).$

\item  Class III transitions: these decays are caused by the interference of color singlet and color neutral currents and consists both color favored and color suppressed diagrams \textit{i.e.} the amplitudes $a_{1} $ and  $a_{2} $ interfere.
\end{enumerate}

For numerical calculations, we follow the convention of taking $N_{c} =3 $ to fix the QCD coefficients $a_{1} $ and $a_{2} $, where  we use \cite{35}:

   \begin{center}
 $c_{1} (\mu ) = 1.12 ,~c_{2} (\mu )=-0.26 $ at $ \mu \approx m_{b}^{2}$.
\end{center}
A detailed analysis regarding $N_c$ counting and role of color-octet current operators is available in \cite{34}. It may be noted that $N_c$, number of color degrees of freedom, may be treated as a phenomenological parameter in weak meson decays, which account for non-factorizable contributions. It implies that the effective expansion parameter is something like, $1/(4\pi)N_c,~ 1/N_c^2...$ or non-leading $1/N_c$ terms are suppressed  by some reason \cite{35}. In order to study the variation in decay rates and branching ratios, we effectively vary the parameter $N_c$ from 3 to 10. The obtained results are thus presented as an average with uncertainties between branching  ratios at $N_c = 3$ to $N_c= 10$.
Taking in to account the constructive interference observed for $B$ meson decays involving both the
color favored and color suppressed diagrams \cite{35}. We use the ratio $a_{2}/a_1$ to be positive in the present calculations.

\subsection{Decay Widths}

Like vector meson $(V)$, axial-vector meson $(A)$ also carry spin degrees of freedom, therefore, the decay rate \cite{31} of $B_c \to VA$ is composed of three independent helicity amplitudes  $H_{0}$, $H_{+1} $ and $H_{-1} $, which is given by
\begin{equation}\Gamma (B_c \to MA)=\frac{p_{c}}{8\pi \, m_{B_{c}}^{2} } (\left|H_{0} \right|^{2} +\left|H_{+1} \right|^{2} +\left|H_{-1} \right|^{2} ),\end{equation}
 where $p_{c} $ is the magnitude of the three-momentum of a final-state particle in the rest frame of $B_{c} $ meson and $M$ = $V$ or $A$. Helicity amplitudes  $H_{0}$, $H_{+1}$ and $H_{-1} $ are defined in terms of the coefficients \textit{a}, \textit{b}, and \textit{c} as follows:
\begin{equation} H_{\pm 1} = a \pm  c (x^{2} -1)^{1/2}, ~ ~ H_{0} = -a x - b (x^{2} -1), \end{equation}
where
\begin{equation} x=\frac{m_{B_{c}}^{2} -m_{M}^{2} -m_{A}^{2} }{2m_{A } m_{M} } ,\end{equation}
such that
\begin{equation} A(B \to MA) \equiv \vp^{*}_{M \mu} \vp^{*}_{A \nu} [a g^{\mu \nu} + b k_{B_c}^{\mu} k_{B_c}^{\nu} + ic\epsilon^{\mu\nu\alpha\beta}k_{B_{c}\alpha} k_{M \beta}  ].\end{equation}

The coefficient \textit{a}, \textit{b} and \textit{c} describe the \textit{s-}, \textit{d-} and \textit{p-} wave contributions, respectively. $m_M$ and $m_{A} $ denotes masses of respective mesons.

\section{Decay Constants}

The decay constants for axial-vector mesons are defined by the matrix elements given in the previous section. It may be pointed out that  the axial-vector meson states are represented by $3 \times 3$ matrix and they transform under the charge conjugation \cite{30} as  \ben
 M_a^b(^3P_1) \to M_b^a(^3P_1), \qquad M_a^b(^1P_1) \to
 -M_b^a(^1P_1),~~~(a=1,2,3).
 \en
 Since the weak axial-vector current transfers as
$(A_\mu)_a^b\to (A_\mu)_b^a$ under charge conjugation, the decay constant of the $^1P_1$ meson should vanish in the SU(3) flavor limit \cite{30}.
Experimental information based on $\tau$ decays gives decay constant $f_{K_{1}} (1270) =0.175 \pm 0.019$ GeV \cite{20,31}, while decay constant for $\underline{K}_{1} (1.400) $ can be obtained from relation $f_{\underline{K}_{1}} (1.400)/f_{K_{1}} (1.270) =\cot \theta _{1} $ $i.e.$
$f_{\underline{K}_{1}} (1.400) =(-0.109\pm 0.12)$ GeV, for $\theta _{1} =-58^{\circ} $ used in the present work \cite{31}.
In case of non-strange axial-vector mesons,  Nardulli and Pham \cite{36} used mixing angle for strange axial vector mesons and SU(3) symmetry to determine $f_{a_{1} } =0.223$ GeV for $\theta_1 = -58^\circ$. Since, $a_1$ and $f_1$ lies in the same nonet we assume $f_{f_{1} } \approx f_{a_{1} } $ under SU(3) symmetry. Due to charge conjugation invariance decay constants for $^1P_1$ nonstrange neutral mesons $b_1^0(1.235)$, $h_1(1.170)$, and $h_1'(1.380)$ vanish. Also, owing to G-parity conservation in the isospin limit decay constant $f_{b_{1}}= 0$.

For decay constants of charmed and strange charmed states, we use $f_{D_{1A}} =-0.127$ GeV, $f_{D_{1A}^{'} } =0.045$ GeV, $f_{D_{s1A}} =-0.121$ GeV, $f_{D_{s1A}^{'} } =0.038$ GeV, and $f_{\chi _{c1} } \approx -0.160$ GeV \cite{34,37}.
\par On the other hand, the decay constants for vector mesons  are relatively trivial, we use
 $f_{\rho } =0.221$ GeV, $f_{K^*} =0.220$ GeV, $f_{D^*} =0.245$ GeV, $f_{D^*_{s} } =0.273$ GeV, $f_{\phi } =0.195$ GeV, $f_{\omega} =0.229$ GeV, and $f_{\J } =0.411$ GeV \cite{2, 15, 31, 37} in numerical calculations.

\section{Form factors}
In this section, we give a short description to calculate $B_{c} \to A$ and $B_{c} \to V$ transition form factors.

\subsection{\pmb{$ B_{c} \to A/A'$} transition form factors}

We use ISGW II Model \cite{27} to calculate $B \to A/A'$ transition form factors. ISGW model is a
non-relativistic constituent quark model \cite{26}, which obtain an exponential $q^2$-dependence of the form factors. It employ variational solutions of the Schrödinger equation based on the harmonic oscillator
wave functions, using the coulomb and linear potential.
In general, the form factors evaluated are considered reliable at $ q^2 = q^2_m$, the maximum
momentum transfer $(m_B-m_X)^2$. The reason being that the form-factor $q^2$-dependence in the
ISGW model is proportional to  $e^{ -(q^2_m-q^2)}$ and hence the form factor decreases
exponentially as a function of $(q^2_m-q^2)$. This has been improved in the ISGW II model \cite{27} in
which the form factor has a more realistic behavior at large $(q^2_m-q^2)$ which is expressed in
terms of a certain polynomial term. In addition to this, the ISGW II model incorporates a number of improvements, such as the heavy quark symmetry constraints, heavy-quark-symmetry-breaking color magnetic interaction, relativistic corrections $etc.$

The form factors have the following simplified expressions in the ISGW II model for $B_{c} \to A/A'$ transitions caused by $b\to c$ quark transition \cite{26,27}:

\ben
l&=&\tilde{m}_{B_{c} } \beta _{B_{c} } [\frac{1}{\mu _{-} } +\frac{m_{c} \tilde{m}_{A} (\tilde{\omega }-1)}{\beta _{B_{c} }^{2} } (\frac{5+\tilde{\omega }}{6m_{q} } -\frac{m_{c} \beta _{B_{c} }^{2} }{2\mu _{-} \beta _{B_{c} A}^{2} } )]\, F_{5}^{(l)} ,\non \\
c_{+} +c_{-} &=&-\frac{\tilde{m}_{A} }{2m_{B_{c} } \beta _{B_{c} }^{} } \left(1-\frac{m_{c}^{2} \beta _{B_{c} }^{2} }{2m_{A} \mu _{-} \beta _{B_{c} A}^{2} } \right)F^{(c_{+} +c_{-} )} ,\non \\
c_{+} -c_{-} &=&-\frac{\tilde{m}_{A} }{2m_{B_{c} } \beta _{B_{c} }^{} } \left(\frac{\tilde{\omega }+2}{3} -\frac{m_{c}^{2} \beta _{B_{c} }^{2} }{2m_{A} \mu _{-} \beta _{B_{c} A}^{2} } \right)F^{(c_{+} -c_{-} )} ,\non \\
q'&=&\frac {m_c}{2 \tilde{m}_{A} \beta _{B_{c} }} (\frac{5+\tilde{\omega }}{6m_{q} }) F_{5}^{(q)},
\en

\ben
r&=&\frac{\tilde{m}_{B_{c} } \beta _{B_{c} } }{\sqrt{2} } [\frac{1}{\mu _{+} } +\frac{\tilde{m}_{A} }{3\beta _{B_{c} }^{2} } (\tilde{\omega }-1)^{2} ]\, F_{5}^{(r)} ,\non \\
s_{+} +s_{-} &=&\frac{m_{c} }{\sqrt{2} \tilde{m}_{B_{c} } \beta _{B_{c} }^{} } \left(\frac{m_{c} \beta _{B_{c} }^{2} }{2\mu _{+} \beta _{B_{c} A}^{2} } \right)F^{(s_{+} +s_{-} )} ,\non \\
   s_{+} -s_{-} &=&\frac{1}{\sqrt{2} \beta _{B_{c} }^{} } \left(\frac{4-\tilde{\omega }}{3} -\frac{m_{c}^{2} \beta _{B_{c} }^{2} }{2\tilde{m}_{A} \mu _{+} \beta _{B_{c} A}^{2} } \right)F^{(s_{+} -s_{-} )} ,\non \\
v&=&[\frac{\tilde m_{B_{c}}\beta_{B_{c}}}{4 \sqrt 2 m_c \tilde m_A}+ \frac {(\tilde \omega-1)m_c}{6\sqrt 2 \tilde m_{A}\beta_{B_{c}}}]F_{5}^{(q)},
   \en
where

\ben
\mu _{\pm } =(\frac{1}{m_{c} } +\frac{1}{m_{b} } )^{-1} ,
\en
 $t(\equiv q^{2} )$ dependence is given by

\ben
\tilde{\omega }-1=\frac{t_{m} -t}{2\bar{m}_{B_{c} } \bar{m}_{A} }.
\en
and
\ben
 F_{5}^{(l)} &=& F_{5}^{(r)} = F_{5} (\frac{\bar{m}_{B_{c} } }{\tilde{m}_{B_{c} } } )^{1/2} (\frac{\bar{m}_{A} }{\tilde{m}_{A} } )^{1/2} , \non \\ F_{5}^{(c_{+} +c_{-} )} &=& F_{5}^{(s_{+} +s_{-} )} = F_{5} (\frac{\bar{m}_{B_{c} } }{\tilde{m}_{B_{c} } } )^{-3/2 } (\frac{\bar{m}_{A} }{\tilde{m}_{A} } )^{1/2} , \non \\ F_{5}^{(c_{+} -c_{-} )} &=& F_{5}^{(s_{+} -s_{-} )} =F_{5} (\frac{\bar{m}_{B_{c} } }{\tilde{m}_{B_{c} } } )^{-1/2} (\frac{\bar{m}_{A} }{\tilde{m}_{A} } )^{-1/2 }, \non \\
  F_{5}^{(q')} &=& F_{5}^{(v)} = F_{5} (\frac{\bar{m}_{B_{c} } }{\tilde{m}_{B_{c} } } )^{-1/2} (\frac{\bar{m}_{A} }{\tilde{m}_{A} } )^{-1/2} .
\en

The function $F_5$ is given by

\ben
F_{5} = \left(\frac{\tilde{m}_{A} }{\tilde{m}_{B_{c} } } \right)^{{\raise0.7ex\hbox{$ 1 $}\!\mathord{\left/ {\vphantom {1 2}} \right. \kern-\nulldelimiterspace}\!\lower0.7ex\hbox{$ 2 $}} } \left(\frac{\beta _{B_{c} } \beta _{A} }{\beta _{B_{c} A} } \right)^{{\raise0.7ex\hbox{$ 5 $}\!\mathord{\left/ {\vphantom {5 2}} \right. \kern-\nulldelimiterspace}\!\lower0.7ex\hbox{$ 2 $}} } \left[1+\frac{1}{18} \chi ^{2} (t_{m} -t)\right]^{-3} ,
\en
with

\ben
\chi ^{2}  = \frac{3}{4m_{b} m_{c} } +\frac{3m_{c}^{2} }{2\bar{m}_{B_{c} } \bar{m}_{A} \beta _{B_{c} A}^{2} } +\frac{1}{\bar{m}_{B_{c} } \bar{m}_{A} } \left(\frac{16}{33-2n_{f} } \right)\ln [\frac{\alpha _{S} (\mu _{QM} )}{\alpha _{S} (m_{c} )} ],
\en
and

\ben
\beta _{B_{c} A}^{2} =\frac{1}{2} \, \, \left(\beta _{B_{c} }^{2} +\beta _{A}^{2} \right).
\en
 $\tilde{m}$ is the sum of the mesons constituent quarks masses, $\bar{m}$ is the hyperfine averaged physical masses, $n_f$ is the number of active flavors, which is taken to be five in the present case, $t_{m} =(m_{B_{c} } -m_{A} )^{2} $ is the maximum momentum transfer and $\mu _{QM} $ is the quark model scale. The values of parameter $\beta $ for different \textit{s}-wave and \textit{p}-wave mesons \cite{26,27} are given in the Table I. We use the following quark masses

 $$ m_u = m_d = 0.31 \pm 0.04,~ m_s = 0.49 \pm 0.04, ~m_c = 1.7 \pm 0.04,~  \textrm {and} ~ m_b = 5.0 \pm 0.04,$$
to calculate the form factors for \textbf{$B_{c} \to A/A'$} transitions which are given in Tables II and III. It may be pointed out that the form factors are sensitive to the choice of quark masses. The variation in quark masses, particularly light quark sector, may lead to uncertainties in the form factors therefore we allowed certain range based on literature \cite{38}. These uncertainties in the form factors are shown in Tables II and III. 

\subsection{\pmb{$B_{c} \to V$} transition form factors}

	For $B_c \to V$ transition form factors we use our previous work \cite{18} based on BSW framework \cite{28}, in which, one of the authors  investigated the possible flavor
dependence in $B_c \to P/V$ form factors and consequently in $B_c \to PP/PV$ decay widths. It may be noted that BSW model \cite{28} the form factors depend upon  the average transverse quark momentum inside a meson $\omega$, which is fixed in the model to 0.40 GeV. However, it has been pointed out that $\omega$ being a dimensional quantity, may show flavor dependence. Therefore, it may not be justified to take the same $\omega$ for all the mesons. Following the analysis described
in \cite{18}, we estimate $\omega$ for different mesons from $|\psi(0)|^2$, $i.e.$ square of the wave function at the origin obtained from
the hyperfine splitting term for the meson masses, which
in turn fixes quark masses (in GeV) to be $ m_u = m_d = 0.31 \pm 0.04,
~m_s = 0.49 \pm 0.04,~ m_c = 1.7 \pm 0.04,$ and $m_b = 5.0 \pm 0.04$ for $\alpha_s(m_b) = 0.19,
~\alpha_s(m_c) = 0.25$, and $\alpha_s = 0.48$ (for light flavors \textit{u}, \textit{d} and \textit{s}). Here also, variation in $\alpha_s$ may lead to uncertainty in quark masses \cite{38} and consequently in form factors. For further details we refer the interested reader to \cite{18}.
We find that all of the form factors get significantly
enhanced due to flavor dependence of $\omega$. The obtained form factors along with corresponding uncertainties due to variation in quark masses are shown in Table IV.

It may also be noted that consistency with the Heavy Quark Symmetry (HQS) requires certain form factors such as $F_1, A_0, A_2 $ and \textit{V} to have dipole $q^2$-dependence \cite{28}. Therefore, we use the following $q^2$-dependence for different form factors:
$$A_{0} (q^{2} )=\frac{A_{0} (0)}{(1-\frac{q^{2} }{m_{P}^{2}} )^{2} } , ~ ~  A_{1} (q^{2} )=\frac{A_{1}(0)}{(1-\frac{q^{2} }{m_{A}^{2}} )^{} },$$
$$A_{2} (q^{2} )=\frac{A_{2}(0)}{(1-\frac{q^{2} }{m_{A}^{2}} )^{2} }~ ~ ~  \textrm{and}  ~~V(q^{2} )=\frac{V(0)}{(1-\frac{q^{2} }{m_{V}^{2}} )^{2} }, $$
with appropriate pole masses $m_{i}$.

\section{Results and Discussions}

Using the decay constants and form factors described in Section IV and V, respectively, we predict the branching ratios of $B_c \to VA$ and $B_c \to AA$ decays in CKM favored and CKM suppressed modes.
\subsection{\pmb{$B_c \to VA$} Decays}
The Branching ratios for $B_c$ decaying to a vector and an axial-vector meson in the final state for CKM favored and CKM suppressed modes are given in column 2 of Tables V-X. We also give the helicity amplitudes of corresponding decay channels in columns 3, 4 and 5 of respective Tables V-X. We observe the following:

\subsubsection*{For CKM favored modes}

\begin{enumerate}
\item The branching ratios for dominant decays in Cabibbo enhanced
($\Delta b= 1,~\Delta C= 1, ~\Delta S=0$) mode are: Br($B_c^- \to \J a_1^-$) =
$4.14 \pm 0.26 \pm 0.05 \t 10^{-3}$; Br($B_c^- \to \rho^- \chi_{c1}$) = $1.47 \pm 0.15 \pm 0.01 \t 10^{-3}$; Br($B_c^- \to \rho^- h_{c1}$) = $1.24 \pm 0.08 \pm 0.01 \t 10^{-3}$. The next order branching ratio is Br($B_c^- \to D^{*0} D_1^-$) = $2.92 \pm 0.84^ {+0.52}_{-0.28} \t 10^{-5}$. We wish to remark here that the first quoted uncertainty in branching ratios is due to effective variation of parameter $N_c$ and the second uncertainty is caused by variation of quark masses in the form factors. The same has been followed throughout the presentation of results including Tables V-X. The branching ratios of the remaining decays are of the order of magnitude $(10^{-6}-10^{-7})$ except for $B_c^- \to \J b_1^-$ decay which is $\mathcal{O} (10^{-8})$.
\item The dominant decay channels in Cabibbo favored ($\Delta b= 1,~\Delta C= 0, ~\Delta S=-1$) mode are those which consist one $\bar{c}c$-meson in the final state \textit{i.e.} Br($B_c^- \to \J D_{s1}^{-}$) = $2.35 \pm 0.25 \pm 0.01 \t 10^{-3}$; ($B_c^- \to D_s^{*-} \chi_{c1}$) = $1.08 \pm 0.08 \pm 0.01 \t 10^{-3}$; Br($B_c^- \to D_s^{*-} h_{c1}$) = $8.11 \pm 0.48 \pm 0.12 \t 10^{-4}$ and Br($B_c^- \to \J \underline{D}_{s1}^{-}$) = $6.33 \pm 0.49 \pm 0.06 \t 10^{-4}$. The rest of the decay modes remains suppressed with branching ratios of $\mathcal{O} (10^{-7} \sim 10^{-11})$.
\item It may be noted that the branching ratios for $B_c \to VA$ decays are higher for axial-vectors $A(^3P_1)$ in the final state as compared to $A(^1P1)$  with same quark content except for strange axial meson emitting decays, which are roughly of the same order.
\item We find that longitudinal helicity amplitudes are higher in magnitude for all the decay modes.
\end{enumerate}

\subsubsection*{For CKM suppressed modes}

\begin{enumerate}
\item It is interesting to note that branching ratios for CKM suppressed mode ($\Delta b= 1,~\Delta C= 1, ~\Delta S=-1$) are of the order ($10^{-4}\sim 10^{-5}$).The dominant decays are:  Br($B_c^- \to \J \underline{K}_{1}^{-}$) = $2.36 \pm 0.14 \pm 0.03 \t 10^{-4}$; ($B_c^- \to \J K_{1}^{-}$) = $1.49 \pm 0.09 \pm 0.02 \t 10^{-4}$; Br($B_c^- \to K^{*-} \chi _{c1}$) = $7.07 \pm 0.43 \pm 0.0.04 \t 10^{-5}$ and Br($B_c^- \to K^{*-} h_{c1}$) = $6.18 \pm 0.37 \pm 0.06 \t 10^{-5}$. The next order decay has Br($B_c^- \to D^{*0} D_{s1}^{-}$) = $2.21\pm 0.63 \pm 0.12\t 10^{-6}$. Branching ratios of the other decay modes are of $\mathcal{O} (10^{-7}-10^{-8})$.
\item Only four decay channels have branching ratios of $\mathcal{O} (10^{-4}-10^{-5})$ in CKM suppressed ($\Delta b= 1,~\Delta C= 1, ~\Delta S=1$) mode \textit{i.e.} Br($B_c^- \to J/\psi D_1^{-}$) = $1.39 \pm 0.14 \pm 0.03 \t 10^{-4}$; Br($B_c^- \to D^{*-} \chi _{c1}$) = $4.95 \pm 0.33 \pm 0.03 \t 10^{-5}$; Br($B_c^- \to D^{*-} h _{c1}$) = $3.88 \pm 0.23 \pm 0.04 \t 10^{-5}$,  Br($B_c^- \to \J \underline{D}_1^{-}$) = $3.45 \pm 0.28 \pm 0.03 \t 10^{-5}$ and Br($B_c^- \to \rho^- \bar{D}_1^{0}$) = $1.01 \pm 0.05 ^{+0.20}_{-0.11} \t 10^{-5}$. Though, branching ratios for $B_c^- \to \rho^- \underline{\bar{D}}_1^{0}$ and $B_c^- \to  \bar{D}^{*0} a_1^-$ decays are of $\mathcal{O} (10^{-6})$, these may also be of experimental interest in near future.
\item In ($\Delta b= 1,~\Delta C= -1, ~\Delta S=-1$) mode, branching ratios for ($B_c^- \to D_s^{*-} \bar{D}_{1}^{0}$), ($B_c^- \to D_s^{*-} \underline{\bar{D}}_{1}^{0}$) and ($B_c^- \to D_{s1}^{-} \bar{D}^{*0}$) decays are  $2.01 \pm 0.13^ {+0.44}_{-0.24} \t 10^{-5}$, $1.71 \pm 0.12 ^ {+0.37}_{-0.19}\t 10^{-6}$ and $ 1.42 \pm 0.22 \pm 0.10 \t 10^{-6}$, respectively. However, branching ratios for ($\Delta b= 1,~\Delta C= -1, ~\Delta S=0$) mode remains highly suppressed.
\item Here also, branching ratios for decays involving $A(^3P_1)$ mesons in the final state are higher than their $A(^1P_1)$ partners for same flavor content. However, for decays involving $K_1$ and $\underline{K}1$ the branching ratios are of same order.
\item The longitudinal helicity amplitudes for the CKM suppressed decays show same trend as observed in CKM favored modes.
\end{enumerate}

\subsection{$\pmb{B_c \to AA}$ Decays}

The calculated branching ratios for $B_c$ decaying to two axial-vector mesons in the final state for CKM favored and CKM suppressed modes are given in column 2 of Tables XI-XVI. The corresponding helicity amplitudes of decay channels are presented in columns 3, 4 and 5 of Tables XI-XVI. Here also, the uncertainties in the obtained results caused by $N_c$ variation and quark mass variation in the form factors, respectively, are given in Tables XI-XVI. We made the following observations:

\subsubsection*{For CKM favored modes}

\begin{enumerate}
\item The branching ratios for $ B_c \to AA$ decays are smaller than those for $B_c \to VA$ decays by an order of magnitude in corresponding CKM modes.
\item The dominant decays in Cabibbo enhanced
($\Delta b= 1,~\Delta C= 1, ~\Delta S=0$) mode have branching ratios: Br($B_c^- \to \chi_{c1} a_1^-$) = $1.81 \pm 0.11 \pm 0.01 \t 10^{-4}$; Br($B_c^- \to h_{c1} a_1^-$) = $1.36 \pm 0.08 \pm 0.10 \t 10^{-4}$. Branching ratios of $B_c^- \to D_1^- D_1^0$ and $B_c^- \to D_1^- \underline{D}_1^0$ decays are of the order of $10^{-6}$. The order of magnitude for branching ratios of the remaining decays ranges from $(10^{-7})\sim (10^{-10})$ in this mode.
\item In Cabibbo favored ($\Delta b= 1,~\Delta C= 0, ~\Delta S=-1$) mode, the dominant decay channels are:   Br($B_c^- \to D_{s1}^{-} \chi_{c1}$) = $1.22 \pm 0.15 \pm 0.05 \t 10^{-4}$; Br($B_c^- \to h_{c1} D_{s1}^{-}$) = $3.15 \pm 0.19 \pm 0.18\t 10^{-5}$; Br($B_c^- \to \underline{D}_{s1}^{-} \chi_{c1}$) = $2.88 \pm 0.27\pm 0.06 \t 10^{-5}$ and Br($B_c^- \to h_{c1} \underline{D}_{s1}^{-} $) = $1.77 \pm 0.07 \pm 0.06 \t 10^{-5}$. The  remaining decay modes are suppressed with branching ratios of $\mathcal{O} (10^{-7} \sim 10^{-11})$.
\item In the present analysis, we observe that magnitude of longitudinal helicity amplitude are higher for all the decay modes except for decays involving $c\bar{c}$ meson in the final state. In such decays transverse helicity amplitude $H_-$ has larger magnitude. 
\end{enumerate}

\subsubsection*{For CKM suppressed modes}

\begin{enumerate}
\item In CKM suppressed mode ($\Delta b= 1,~\Delta C= 1, ~\Delta S=-1$) the highest order of magnitude for branching ratios of dominant decays is $\sim(10^{-5}-10^{-6})$ \textit{i.e.} Br($B_c^- \to \chi_{c1} \underline{K}_{1}^{-}$) = $1.18 \pm 0.07 \pm 0.01 \t 10^{-5}$; Br($B_c^- \to \chi_{c1} K_{1}^{-}$) = $6.76 \pm 0.40 \pm 0.04 \t 10^{-6}$;  Br($B_c^- \to h_{c1} \underline{K}_{1}^{-}$) = $6.63 \pm 0.40 \pm 0.45 \t 10^{-6}$ and Br($B_c^- \to h_{c1} K_{1}^{-}$) = $4.99 \pm 0.30 \pm 0.34 \t 10^{-6}$. The next order decays have branching ratios of the other decay modes are of $\mathcal{O} (10^{-7} \sim 10^{-8})$.
\item Likewise ($\Delta C= 1, ~\Delta S=-1$), the dominant decay channels have branching ratios of $\mathcal{O} (10^{-6})$ in CKM suppressed ($\Delta b= 1,~\Delta C= 1, ~\Delta S=1$) mode \textit{i.e.} Br($B_c^- \to D_1^{-} \chi _{c1}$) = $7.48 \pm 0.92 \pm 0.50\t 10^{-6}$; Br($B_c^- \to \bar{D}_1^{0} a_1^-$) = $3.41 \pm 0.20 ^ {+1.30}_{-0.77}\t 10^{-6}$; Br($B_c^- \to \underline{D}_1^{-} \chi _{c1}$) = $1.60\pm 0.15 \pm 0.04 \t 10^{-6}$; and Br($B_c^- \to D_1^{-} h_{c1}$)= $1.79 \pm 0.11 \pm 0.01 \t 10^{-6}$. However, branching ratios for $B_c^- \to a_1^- \underline{\bar{D}}_1^{0}$ and $B_c^- \to \underline{D}_1^{-} h_{c1}$ decays are of $\mathcal{O} (10^{-7})$.
\item Decay channels in CKM suppressed ($\Delta b= 1,~\Delta C= -1, ~\Delta S=-1$) and  ($\Delta b= 1,~\Delta C= -1, ~\Delta S=0$) modes remain highly suppressed with Br($B_c^- \to  \underline{D}_{s1}^{-}\underline{\bar{D}}_{1}^0$) $= 1.76 \pm 0.15 ^ {+0.55}_{-0.34} \t 10^{-6}$. $B_c^- \to \underline{\bar{D}}_1^{0}D_{s1}^-$ and $B_c^- \to \underline{\bar{D}}_{s1}^{-} \bar{D}_{1}^0$ decays have branching ratios $\mathcal{O}(10^{-7})$.
\item As noticed in the previous case, longitudinal helicity amplitudes have larger magnitude in comparison to transverse components for most of the decay channels. However, decay channels involving $c\bar{c}$ meson in the final state show transverse $H_-$ component dominance.
  
\end{enumerate}
It may also be noted that effective variation in $N_c$ leads to the change in amplitude and hence, branching ratios of these decays. The branching ratios of color favored class I decays show $\sim$ 6\% variation in the central value and color suppressed class II decays show variation of $\sim$ 30\%. However, class III decays involving both color favored and color suppressed diagrams show a variation from 7\% to 15\%.

We wish to emphasize that with remarkable improvements in experiment and sophisticated instrumentation branching ratios of the order of $(10^{-6})$ could be measured precisely \cite{39} at LHC, LHC-b and Super-B factories in near future. Therefore, it may provide the necessary  information for phenomenological study of $B_c$ meson physics.

Since, there is no experimental information available at present for such decays, we compare our results with other theoretical works (see Table XVII). There are several theoretical models like Bethe-Salpeter approach (BSA) \cite{25}, Relativistic Quark Model (RQM) \cite{13, 23},  Non Relativistic Quark Model (NRQM) \cite{15} $etc.$ which give their predictions for $B_c \to VA$ decays with charmonium in the final state. We find that results given by different models are comparable with some exceptions. We have used $a_1 = 1.12$ to obtain branching ratios for these models in Table of comparison. It may be noted that H.F. Fu \textit{et al.} \cite{24} also predict branching ratios of few decay modes namely $B_c^- \to h_{c1} D^{*-}/\chi_{c1} D^{*-}/J/\psi D_{s1}^-/J/\psi  \underline{D}_{s1}^-$. Their predictions are lager than our results by an order of magnitude except for $B_c^- \to  \chi_{c1} D^{*-}$ which is comparable to our prediction. In addition to these, H.F. Fu \textit{et al.} \cite{24} predict branching ratios of  $B_c^- \to D_{s1}^- \phi^0/ \underline{D}_{s1}^- \phi^0/ \underline{D}_{s1}^-K^{*0} $ decays based on contributions from penguin diagrams which we ignore in the present analysis. We wish remark here that for $B_c \to AA$ decays, theoretical predictions for only four decay channels are available for comparison \textit{i.e.} $B_c^- \to h_{c1} D_{s1}^-/ h_{c1} \underline{D}_{s1}^-/\chi_{c1} D_{s1}^-/ \chi_{c1} \underline{D}_{s1}^-$ \cite{24}. Here also, branching ratios predicted in present work are small as compared to results given by \cite{24}.

\section{Summary and Conclusions}

In the present work we have calculated $B_c \to A$ transition form factors using ISGW II model framework. Consequently, we have predicted branching ratios of $B_c \to VA/AA$ decays. We have used flavor dependent $B_c \to V$ transition form factors in BSW Model framework. Also, we have calculated the helicity components corresponding to different polarization amplitudes in $B_c \to VA/AA$ decays. We draw the following conclusions:
\begin{enumerate}
\item In case of $B_c \to VA$ mode, CKM enhanced ($\Delta b= 1,~\Delta C= 1, ~\Delta S=0$) dominant decays are $B_c^- \to \J a_1^-$, $B_c^- \to \rho^- \chi_{c1}$ and $B_c^- \to \rho^- h_{c1}$, while the dominant decays in ($\Delta b= 1,~\Delta C= 0, ~\Delta S=-1$) are $B_c^- \to D_s^{*-} \chi_{c1}$, $B_c^- \to D_s^{*-} h_{c1}$, $B_c^- \to \J D_{s1}^{-}$, $B_c^- \to \J \underline{D}_{s1}^{-}$. Their branching ratios range from $10^{-3}-10^{-11}$.
\item Branching ratios of CKM enhanced modes in case of $B_c \to AA$ decays are smaller by an order of magnitude in comparison to those in $B_c \to VA$ decays. The dominant decays are: $B_c^- \to \chi_{c1} a_1^-$,  $B_c^- \to h_{c1} a_1^-$ and $ B_c^- \to D_{s1}^- \chi_{c1}$. Here, also the branching ratios range from $10^{-4}-10^{-10}$.
\item In CKM suppressed modes, the branching ratios are further small by an order of magnitude for both $B_c \to VA$ and $B_c \to AA$ decays. The branching ratios for the dominant decays $B_c^- \to \J K_{1}^{-}$, $B_c^- \to \J \underline{K}_{1}^{-}$ and $B_c^- \to \J D_{1}^{-}$ are of the order of magnitude ($10^{-4}$).
\item In general, branching ratios of $B_c \to VA$ decays involving axial-vector $A(^3P_1)$ in the final state are larger in comparison to the decays involving axial-vector $A(^1P_1)$ in final state.
\item For most of the decays, magnitude of helicity component for longitudinal polarization amplitude is larger in comparison to the transverse amplitudes. However, in $B_c \to AA$ decays transverse polarization amplitude dominance has been observed for channels involving $c\bar{c}$ meson in final state. 
\end{enumerate}
Since, LHC and LHC-b are expected to accumulate data for more than $10^{10}$ $B_c$ events per year, we hope that predicted BRs would be measured soon in these experiments.

\begin{acknowledgments}
The work was supported by the National Research Foundation of Korea (NRF)
grant funded by Korea government of the Ministry of Education, Science and
Technology (MEST) (No. 2011-0017430) and (No. 2011-0020333).
\end{acknowledgments}

\newpage

\begin{table}
\captionof{table} {The values of $\beta$ parameter  for $s$-wave and $p$-wave mesons in the ISGW II quark model.}
\label{t1}
\begin{tabular}{ c c c c c c c c c c } \toprule
Quark content  & $u\bar{d}$ & $u\bar{s}$ & $s\bar{s}$ & $c\bar{u}$ & $c\bar{s}$ & $u\bar{b}$ & $s\bar{b}$ & $c\bar{c}$ & $b\bar{c}$   \\ \hline
$\beta _{s} $(GeV)  & 0.41  & 0.44  & 0.53 & 0.45  & 0.56 & 0.43  & 0.54 & 0.88 & 0.92 \\ \hline
$\beta _{p} $ (GeV)  & 0.28  & 0.30  & 0.33 & 0.33  & 0.38 & 0.35  & 0.41 & 0.52 & 0.60 \\ \hline
\end{tabular}
\end{table}

\begin{subtables}
\begin{table}
\captionof{table} {$B_{c} \to A$ transition form factors at $q^2 _{max.} $ in the ISGW II quark model.}
\label{t2a}
\begin{tabular}{ c c c c c c }
\toprule
Modes & Transition & $l$ & $c_+$ & $c_-$ & $q'$ \\
\hline
\multirow {2}{*}{$\Delta b =1, \Delta C = 0, \Delta S = -1$} & $B_{c} \to D_{1}$  &$ -3.529 ^{+0.504}_{-0.430} $ & $-0.048 \pm 0.001 $ & $-0.006 \pm 0.00 $ & $-0.074\pm 0.002$   \\ \cline{2-6}
  & $B_{c} \to D_{s1}$ &$-2.860\pm 0.258 $ & $-0.061\pm 0.001 $ &  $-0.006\pm 0.001 $ & $-0.095\pm 0.002 $  \\ \hline
$ \Delta b =1, \Delta C = 1, \Delta S = 0$ & $B_{c} \to \chi_{c1}(c \bar {c})$ &$-1.182 \pm 0.038 $ & $-0.103 \pm 0.003 $ & $-0.006 \pm 0.001 $ & $-0.130\pm 0.003$  \\
\hline
\multirow {3}{*}{$\Delta b =1, \Delta C = 0, \Delta S = -1$}
& $B_{c}^{} \to a_1$ &$ -0.243 \pm 0.008 $ & $-0.036\pm 0.001 $ & $0.015\pm 0.001 $ & $-0.074\pm 0.002   $\\ \cline{2-6}
& $B_{c}^{} \to f_1$  &$-0.242\pm 0.008 $ & $-0.036\pm 0.001 $ & $0.015\pm 0.001 $ & $-0.074\pm 0.002$   \\ \cline{2-6}
 & $B_{c}^{-} \to f_{1}^{'} $ &$-0.363\pm 0.010 $ & $-0.049\pm 0.002 $ & $0.018\pm 0.001 $ & $-0.095\pm 0.002 $ \\ \hline
\bigskip
\bigskip

\bigskip
\bigskip
\end{tabular}

\captionof{table} {$B_{c} \to A$  transition form factors (in BSW model type notations) at $q^2_{max.} $ in the ISGW II quark model.}
\label{t2b}
\begin{tabular}{c c c c c c} \toprule
Modes & Transition & $A$ & $V_1$& $V_2$& $V_0$ \\ \hline
\multirow {2}{*}{$\Delta b =1, \Delta C = 0, \Delta S = -1$} & $B_{c}^{} \to D_1$ &$ 0.646 \pm 0.15$ & $-0.406^{+0.63}_{-0.50} $ & $0.421^{+0.004}_{-0.003} $ & $-1.081\pm ^{+0.112}_{-0.097}$ \\ \cline{2-6}
 & $B_{c}^{-} \to D_{s1}^{} $ &$ 0.829\pm 0.020 $ & $-0.326\pm 0.029 $ & $0.535\pm 0.011 $ & $-1.00\pm 0.044$\\ \hline
$\Delta$\textit{b }=1, $\Delta$\textit{C = }1, $\Delta$\textit{S = }0 & $B_{c}^{} \to \chi _{c1} (c\bar{c})$ &$ 1.273 \pm 0.030 $ & $-0.120 \pm 0.005 $ & $1.008\pm 0.032 $ & $-0.572\pm 0.007 $\\ \hline
\multirow {3}{*}{$\Delta b =1, \Delta C = 0, \Delta S = -1$}
& $B_{c}^{} \to a_1$ & $0.553 \pm 0.013 $ & $-0.032\pm 0.001 $ & $0.270\pm 0.010 $ & $-0.495\pm 0.009 $\\ \cline{2-6}
& $B_{c}^{} \to f_1$ & $0.558 \pm 0.013 $ & $-0.032\pm 0.001 $ & $0.272\pm 0.009 $ & $-0.476\pm 0.009 $\\ \cline{2-6}
 & $B_{c}^{-} \to f_{1}^{'} $ &$ 0.733\pm 0.017 $ & $-0.047\pm 0.001 $ & $0.378\pm 0.013 $ & $-0.626\pm 0.011$ \\ \hline
 \bigskip
\bigskip
\end{tabular}
\end{table}
\end{subtables}
\newpage

\begin{subtables}
\begin{table}
\captionof{table} {$B_{c} \to A'$ transition form factors at $q^2 _{max.} $ in the ISGW II quark model.}
\label{t3a}
\begin{tabular}{ c c c c c c } \toprule
Modes & Transition & $r$ & $s_+$ & $s_-$ & $v$ \\ \hline
\multirow {2}{*}{$\Delta b =1, \Delta C = 0, \Delta S =-1$} &  $B\to \underline{D}_{1}^{} $ &$ 2.825 ^{+0.355}_{-0.306} $ & $0.083 \pm 0.000 $ & $-0.055 ^{+0.005}_{-0.003} $ & $0.057^{+0.009}_{-0.006}$ \\ \cline{2-6}
 & $B_{c}^{} \to \underline{D}_{s1}^{} $ &$ 2.464 \pm 0.193 $ & $0.102 \pm 0.001 $ & $-0.060 \pm 0.002 $ & $0.046\pm 0.003$ \\ \hline
$\Delta$\textit{b }=1, $\Delta$\textit{C = }1, $\Delta$\textit{S = }0 & $B_{c}^{} \to h_{c1} (c\bar{c})$ &$ 1.674 \pm 0.044 $ & $0.143 \pm 0.004 $ & $-0.039 \pm 0.001 $ & $0.019\pm 0.001$\\ \hline
\multirow {3}{*}{$\Delta b =1, \Delta C = 0, \Delta S = -1$}
& $B_{c}^{} \to b_1$ & $0.344 \pm 0.007 $ & $0.053\pm 0. 001$ & $-0.028\pm 0.001 $ & $0.010\pm 0.000 $\\ \cline{2-6}
& $B_{c}^{} \to h_1$ & $0.337 \pm 0.007 $ & $0.054\pm 0.001 $ & $-0.029\pm 0.000 $ & $0.011\pm 0.001 $\\ \cline{2-6}
 & $B_{c}^{-} \to h_{1}^{'} $ & $0.512\pm 0.010 $ & $0.074\pm 0.002 $ & $-0.037\pm 0.001 $ & $0.014\pm 0.001$   \\ \hline
 \bigskip
 \bigskip
 \bigskip
\bigskip
\bigskip
\end{tabular}
\captionof{table} {$B_{c} \to A'$  transition form factors (in BSW model type notations) at $q^2_{max.} $ in the ISGW II quark model.}
\label{t3b}
\begin{tabular}{c c c c c c} \toprule
Modes & Transition & $A$ & $V_1$& $V_2$& $V_0$ \\ \hline
\multirow {2}{*}{$\Delta b =1, \Delta C = 0, \Delta S = -1$} & $B_{c}^{} \to \underline{D}_1$ &$ -0.498 ^{+0.067}_{-0.055} $ & $0.324^{+0.042}_{-0.034} $ & $-0.722\pm 0.005 $ & $0.987^{+0.067}_{-0.53}$ \\ \cline{2-6}
 & $B_{c}^{-} \to \underline{D}_{s1}^{} $ & $-0.407\pm 0.033 $ & $0.280\pm 0.022 $ & $-0.898\pm 0.009 $ & $0.995\pm 0.025 $\\ \hline
$\Delta$\textit{b }=1, $\Delta$\textit{C = }1, $\Delta$\textit{S = }0 & $B_{c}^{} \to h _{c1} (c\bar{c})$ &$ -0.183 \pm 0.004 $ & $0.171\pm 0.003 $ & $-1.401\pm 0.032 $ & $0.742\pm 0.008$  \\ \hline
\multirow {3}{*}{$\Delta b =1, \Delta C = 0, \Delta S = -1$}
& $B_{c}^{} \to b_1$ &$ -0.079 \pm 0.002 $ & $0.046\pm 0.001 $ & $-0.401\pm 0.008 $ & $0.677\pm 0.010 $ \\ \cline{2-6}
& $B_{c}^{} \to h_1$ &$ -0.080 \pm 0.002 $ & $0.045\pm 0.001 $ & $-0.402\pm 0.008 $ & $0.702\pm 0.010 $ \\ \cline{2-6}
& $B_{c}^{-} \to h_{1}^{'} $ &$ -0.105\pm 0.002 $ & $0.067\pm 0.001 $ & $-0.566\pm 0.012 $ & $0.866\pm 0.012$ \\ \hline
\end{tabular}
\end{table}
\end{subtables}

\begin{table}
\captionof{table} {$B_{c} \to V$ transition form factors at $q^2 = 0 $ using flavor dependent $\omega$ in BSW model[34].}
\label{t4}
\begin{tabular}{c c c c c c} \toprule
Modes & Transition & $V$ & $A_1$& $A_2$& $A_0$ \\ \hline
\multirow {2}{*}{$\Delta b =1, \Delta C = 0, \Delta S = -1$} & $B_{c} \to D^*$ &$ 0.161 \pm 0.014 $ & $0.094\pm 0.009 $ & $0.108\pm 0.012 $ & $0.079\pm 0.008$ \\ \cline{2-6}
 & $B_{c} \to D_{s}^{*} $ & $0.284\pm 0.09 $ & $0.171\pm 0.008 $ & $0.193\pm 0.010 $ & $0.150\pm 0.008$ \\ \hline
$\Delta$\textit{b }=1, $\Delta$\textit{C = }1, $\Delta$\textit{S = }0 & $B_{c} \to \J (c\bar{c})$ &$ 0.919 \pm 0.002 $ & $0.624\pm 0.008 $ & $0.741\pm 0.020 $ & $0.564\pm 0.001$ \\ \hline
 \multirow {3}{*}{$\Delta b =1, \Delta C = 0, \Delta S = -1$} &  $B_{c} \to \rho$ &$ 0.369 \pm 0.023 $ & $0.577\pm 0.042 $ & $0.624\pm 0.046 $ & $0.410\pm 0.028 $\\ \cline{2-6} &
 $B_{c}  \to \omega$ & $0.272 \pm 0.020 $ & $0.424\pm 0.036 $ & $0.460\pm 0.040 $ & $0.296\pm 0.024 $\\ \cline{2-6}&
 $B_{c}  \to \phi$ & $0.150 \pm 0.017 $ & $0.217\pm 0.026 $ & $0.245\pm 0.029 $ & $0.144\pm 0.017$ \\ \hline
\end{tabular}
\end{table}

\newpage
\begin{table}
\captionof{table} {Branching ratios and helicity amplitudes of $B_{c}  \to  VA$ decays for CKM-favored ($\Delta b=1,\Delta C=0,\Delta S=-1$) mode.}
\footnotesize
\label{t5}
\begin{tabular}{c c c c c} \hline

\multirow{2}{*}{Decays} &  \multirow{2}{*}{Branching Ratios} & \multicolumn{3}{c}{Helicity Amplitudes} \\ \cline{3-5}
 \multicolumn{2}{c}{}& $|H_0|$& $|H_+|$& $|H_-|$ \\  \hline
$B_c^-\J  a_1^-$ & $  4.14 \pm 0.26 \pm 0.05 \t 10^{-3}$ & $1.66 \pm 0.05 \pm 0.01 \t 10^{-1}$ & $ 3.68 \pm 0.11 \pm 0.08 \t 10^{-2}$ & $ 1.04 \pm 0.03 \pm 0.01 \t 10^{-1}$
\\ \hline

$B_c^-\J  b_1^-$ & $  2.90 \pm 0.18 \pm 0.04 \t 10^{-8}$ & $4.38 \pm 0.13 \pm 0.02 \t 10^{-4}$ & $ 9.77 \pm 0.33 \pm 0.23 \t 10^{-5}$ & $ 2.74 \pm 0.07 \pm 0.03 \t 10^{-4}$   \\ \hline

$B_c^- \to  \rho ^- \chi _{c1}$ & $  1.47 \pm 0.15 \pm 0.01 \t 10^{-3}$ & $1.16 \pm 0.04 \pm 0.01 \t 10^{-1}$ & $ 2.70 \pm 0.08 \pm 0.01 \t 10^{-2}$ & $3.83 \pm 0.12 \pm 0.31 \t 10^{-3}$ \\ \hline

$B_c^- \to  \rho ^-  h_{c1}$  & $  1.24 \pm 0.08 \pm 0.01 \t 10^{-3}$ & $ 1.12 \pm 0.04 \pm 0.01 \t 10^{-1}$ & $1.05 \pm 0.03 \pm 0.03 \t 10^{-2}$ & $5.10 \pm 0.15 \pm 0.17 \t 10^{-3}$   \\ \hline

$B_c^- \to  D^{*0} D_1^-$ & $  2.92 \pm 0.84 ^{+0.52}_{-0.28} \t 10^{-5}$ & $1.52 \pm 0.22 ^{+0.11}_{-0.9} \t 10^{-2}$ & $ 5.64 \pm 0.83 ^{+0.61}_{-0.37} \t 10^{-3}$ & $ 2.67 \pm 0.30 ^{+0.40}_{-0.26} \t 10^{-3}$  \\ \hline

$B_c^- \to  D^{*0} \underline{D}_{1}^{-}$ & $  2.43 \pm 0.70 ^{+0.44}_{-0.23} \t 10^{-6}$ & $ 4.38 \pm 0.64 ^{+0.47}_{-0.28} \t 10^{-3}$ & $1.52 \pm 0.22 \pm 0.02 \t 10^{-3}$ & $0.91 \pm 0.14 ^{+2.04}_{-1.60} \t 10^{-3}$
 \\ \hline

$B_c^- \to  D^{*-} D_1^0$ &  $  1.19 \pm 0.34 \pm 0.16 \t 10^{-6}$ & $1.96 \pm 0.29 \pm 0.16 \t 10^{-3}$ & $ 1.35 \pm 0.20 \pm 0.28 \t 10^{-4}$ & $ 2.67 \pm 0.39 \pm 0.22 \t 10^{-3}$
 \\ \hline

$B_c^- \to  D^{*-} \underline{D}_{1}^{0}$  & $  3.71 \pm 1.06 \pm 0.51 \t 10^{-7}$ & $ 1.09 \pm 0.16 \pm 0.93 \t 10^{-3}$ & $7.43 \pm 1.09 \pm 1.55 \t 10^{-5}$ & $1.49 \pm 0.22 \pm 0.12 \t 10^{-3}$ \\ \hline

\end{tabular}
\end{table}
 \begin{table}
\captionof{table} {Branching ratios and helicity amplitudes of $B_{c}  \to  VA$ decays for CKM-favored ($\Delta  b = 1, \Delta C = 0, \Delta S = -1$) mode.}
\label{t6}
\footnotesize
\begin{tabular}{c c c c c} \hline
\multirow{2}{*}{Decays} &  \multirow{2}{*}{Branching Ratios} & \multicolumn{3}{c}{Helicity Amplitudes} \\ \cline{3-5}
 \multicolumn{2}{c}{}& $|H_0|$& $|H_+|$& $|H_-|$ \\  \hline

$B_c^-\J  D_{s1}^-$ & $  2.35 \pm 0.25 \pm 0.01 \t 10^{-3}$ & $1.26 \pm 0.18 \pm 0.01 \t 10^{-1}$ & $ 7.06 \pm 0.49 \pm 0.02 \t 10^{-2}$ & $ 1.04\pm 0.05 \pm 0.00 \t 10^{-1}$
 \\ \hline

$B_c^-\J  \underline{D}_{s1}^{-}$ & $  6.33 \pm 0.49 \pm 0.06 \t 10^{-4}$ & $ 6.18 \pm 0.25 \pm 0.02 \t 10^{-2}$ & $3.73 \pm 0.19 \pm 0.05 \t 10^{-2}$ & $6.57 \pm 0.22 \pm 0.02 \t 10^{-2}$ \\ \hline

$B_c^- \to  D_s^{*-} \chi _{c1}$ & $  1.08 \pm 0.08 \pm 0.01 \t 10^{-3}$ & $9.85 \pm 0.20 \pm 0.02 \t 10^{-2}$ & $ 8.05 \pm 0.25 \pm 0.03 \t 10^{-2}$ & $ 3.10 \pm 1.40 \pm 0.92 \t 10^{-2}$  \\ \hline

$B_c^- \to  D_s^{*-}  h_{c1}$ & $  8.11 \pm 0.48 \pm 0.12 \t 10^{-4}$ & $ 1.02 \pm 0.03 \pm 0.05 \t 10^{-1}$ & $3.97 \pm 0.11 \pm 0.09 \t 10^{-2}$ & $2.61 \pm 0.08 \pm 0.07 \t 10^{-2}$
 \\ \hline

$B_c^- \to  K^{*-} \bar{D}_1^0$ & $  5.29 \pm 0.31 ^{+1.12}_{-0.57} \t 10^{-7}$ & $2.04 \pm 0.06 ^{+0.20}_{-0.12} \t 10^{-3}$ & $ 3.10 \pm 0.09 ^{+0.35}_{-0.21} \t 10^{-4}$ & $ 1.04 \pm 0.03 ^{+0.22}_{-0.14} \t 10^{-4}$ \\ \hline

$B_c^- \to  K^{*-} \underline{\bar{D}}_{1}^{0}$  & $ 6.08 \pm 0.37 ^{+1.35}_{-0.67} \t 10^{-8}$ & $ 6.92 \pm 0.20 ^{+0.74}_{-0.42} \t 10^{-4}$ & $8.57 \pm 0.26 \pm 0.40 \t 10^{-5}$ & $5.12 \pm 0.15 ^{+1.26}_{-0.92} \t 10^{-5}$ \\ \hline

$B_c^- \to  \bar{D}^{*0} K_1^-$ & $  3.92 \pm 0.23 \pm 0.73 \t 10^{-8}$ & $ 4.57 \pm 0.13 \pm 0.45 \t 10^{-4}$ & $1.82 \pm 0.05 \pm 0.39 \t 10^{-5}$ & $3.12 \pm 0.09 \pm 0.29 \t 10^{-4}$ \\ \hline

$B_c^- \to  \bar{D}^{*0} \underline{K}_{1}^{-}$ & $  6.29 \pm 0.36 \pm 1.17 \t 10^{-8}$ & $5.66 \pm 0.14 \pm 0.55 \t 10^{-4}$ & $ 2.72 \pm 0.08 \pm 0.51 \t 10^{-5}$ & $ 4.19 \pm 0.12 \pm 0.39 \t 10^{-4}$
\\ \hline

$B_c^- \to  D_s^{*-} a_1^0$ & $  1.43 \pm 0.41 \pm 0.08 \t 10^{-9}$ & $0.89 \pm 0.13 \pm 0.31 \t 10^{-5}$ & $ 5.36 \pm 0.78 \pm 0.54 \t 10^{-6}$ & $ 5.58 \pm 0.82 \pm 0.21 \t 10^{-5}$  \\ \hline

$B_c^- \to  D_s^{*-} f_1$ &  $ 1.44  \pm 0.41 \pm 0.08 \t 10^{-9}$ & $ 0.89 \pm 0.13 \pm 0.31 \t 10^{-5}$ & $ 5.55 \pm 0.81 \pm 0.54 \t 10^{-5}$& $5.79 \pm 0.85 \pm 0.21 \t 10^{-5}$  \\ \hline

$B_c^- \to  \phi  D_{s1}^-$ & $  6.61 \pm 1.90 \pm 0.30 \t 10^{-9}$ & $2.27 \pm 0.33  \pm 0.06 \t 10^{-4}$ & $ 3.73 \pm 0.55\pm 0.10 \t 10^{-5}$ & $ 0.95 \pm 0.14 \pm 0.07 \t 10^{-5}$ \\ \hline

$B_c^- \to  \phi  \underline{D}_{s1}^{-}$  & $  4.79 \pm 1.38 \pm 0.20 \t 10^{-10}$ & $ 6.08 \pm 0.90 \pm 0.16 \t 10^{-5}$ & $1.32 \pm 0.19 \pm 0.15 \t 10^{-5}$ & $3.68 \pm 0.54 \pm 0.47 \t 10^{-6}$ \\ \hline

$B_c^- \to  \rho ^0 D_{s1}^-$ & $  6.65 \pm 1.94 \pm 0.30 \t 10^{-9}$ & $2.27 \pm 0.33 \pm 0.06 \t 10^{-4}$ & $ 2.79 \pm 0.41 \pm 0.10 \t 10^{-5}$ & $ 7.02 \pm 1.03 \pm 0.72 \t 10^{-6}$ \\ \hline

$B_c^- \to  \rho ^0 \underline{D}_{s1}^{-}$  & $  5.00 \pm 1.44 \pm 0.20 \t 10^{-10}$ & $ 6.22 \pm 0.91 \pm 0.15 \t 10^{-5}$ & $1.00 \pm 0.15 \pm 0.02 \t 10^{-5}$ & $2.79 \pm 0.41 \pm 0.47 \t 10^{-6}$ \\ \hline

$B_c^- \to  \omega  D_{s1}^-$ & $  4.15 \pm 1.19 \pm 0.18 \t 10^{-10}$ & $5.67 \pm 0.83 \pm 0.15 \t 10^{-5}$ & $ 7.08 \pm 1.04 \pm 0.25 \t 10^{-6}$ & $ 1.78 \pm 0.26 \pm 0.18 \t 10^{-6}$ \\ \hline

$B_c^- \to  \omega  \underline{D}_{s1}^{-}$  & $ 3.12 \pm 0.90 \pm 0.12 \t 10^{-11}$ & $ 1.55 \pm 0.23 \pm 0.04 \t 10^{-5}$ & $2.53 \pm0.37 \pm 0.04 \t 10^{-6}$ & $7.08 \pm 1.04 \pm 1.10 \t 10^{-7}$  \\ \hline

\end{tabular}
\end{table}

\begin{table}
\captionof{table} {Branching ratios and helicity amplitudes of $B_{c}  \to  VA$ decays for CKM-suppressed ($\Delta  b = 1, \Delta C = 1, \Delta S = -1$) mode.}
\label{t7}
\footnotesize
\begin{tabular}{c c c c c} \hline
\multirow{2}{*}{Decays} &  \multirow{2}{*}{Branching Ratios} & \multicolumn{3}{c}{Helicity Amplitudes} \\ \cline{3-5}
 \multicolumn{2}{c}{}& $|H_0|$& $|H_+|$& $|H_-|$ \\  \hline
$B_c^-\J  K_1^-$ & $  1.49 \pm 0.09 \pm 0.02 \t 10^{-4}$ & $ 3.12 \pm 0.09 \pm 0.02 \t 10^{-2}$ & $7.22 \pm 0.23 \pm 0.17 \t 10^{-3}$ & $2.03 \pm 0.06 \pm 0.02 \t 10^{-2}$ \\ \hline

$B_c^-\J  \underline{K}_{1}^{-}$ & $  2.36 \pm 0.14 \pm 0.03 \t 10^{-4}$ & $3.81 \pm 0.07 \pm 0.02 \t 10^{-2}$ & $ 1.00 \pm 0.04 \pm 0.02 \t 10^{-2}$ & $ 2.72 \pm 0.08 \pm 0.02 \t 10^{-2}$  \\ \hline
$B_c^- \to  K^{*-} \chi _{c1}$ & $  7.07 \pm 0.43 \pm 0.04 \t 10^{-5}$ & $2.59 \pm 0.08 \pm 0.01 \t 10^{-2}$ & $ 7.10 \pm 0.21 \pm 0.00 \t 10^{-3}$ & $1.01 \pm 0.03 \pm 0.06 \t 10^{-3}$  \\ \hline

$B_c^- \to  K^{*-}  h_{c1}$ & $  6.18 \pm 0.37 \pm 0.06 \t 10^{-5}$ & $ 2.50 \pm 0.07 \pm 0.01 \t 10^{-2}$ & $2.79 \pm 0.08 \pm 0.08 \t 10^{-3}$ & $1.37 \pm 0.06 \pm 0.04 \t 10^{-3}$  \\ \hline

$B_c^- \to  D^{*0} D_{s1}^-$ & $2.21 \pm 0.63 \pm 0.12\t 10^{-6}$ & $ 4.25 \pm 0.62 \pm 0.13 \t 10^{-3}$ & $1.50 \pm 0.22 \pm 0.06 \t 10^{-3}$ & $4.25 \pm 0.62 \pm 0.45 \t 10^{-4}$ \\ \hline

$B_c^- \to  D^{*0} \underline{D}_{s1}^{-}$  & $  1.27 \pm 0.37 \pm 0.05 \t 10^{-7}$ & $ 0.96 \pm 0.14 \pm 0.03 \t 10^{-3}$ & $5.12 \pm 0.76 \pm 0.07 \t 10^{-4}$ & $1.34 \pm 0.20 \pm 0.26 \t 10^{-4}$   \\ \hline

$B_c^- \to  D_s^{*-} D_1^0$ & $ 1.98 \pm 0.57 \pm 0.12 \t 10^{-7}$ & $8.42 \pm 1.23 \pm 0.30 \t 10^{-4}$ & $ 1.03 \pm 0.15 \pm 0.10 \t 10^{-4}$ & $ 1.07 \pm 0.16 \pm 0.04 \t 10^{-3}$  \\ \hline

$B_c^- \to  D_s^{*-} \underline{D}_{1}^{0}$  & $  6.17 \pm 1.77 \pm 0.38 \t 10^{-8}$ & $ 4.70\pm 0.69 \pm 0.17 \t 10^{-4}$ & $5.73 \pm 0.84 \pm 0.58 \t 10^{-5}$ & $5.98 \pm 0.88 \pm 0.22 \t 10^{-4}$
 \\ \hline

\end{tabular}
\end{table}
\begin{table}
\captionof{table} {Branching ratios and helicity amplitudes of $B_{c}  \to  VA$ decays for CKM-suppressed ($\Delta  b = 1, \Delta C = 1, \Delta S = -1$) mode.}
\label{t8}
\footnotesize
 \begin{tabular}{c c c c c} \hline
 \multirow{2}{*}{Decays} &  \multirow{2}{*}{Branching Ratios} & \multicolumn{3}{c}{Helicity Amplitudes} \\ \cline{3-5}
 \multicolumn{2}{c}{}& $|H_0|$& $|H_+|$& $|H_-|$ \\  \hline
$B_c^-\J  D_1^-$ & $  1.39 \pm 0.14 \pm 0.03 \t 10^{-4}$ & $ 3.00 \pm 0.18 \pm 0.04 \t 10^{-2}$ & $1.64 \pm 0.10 \pm 0.02 \t 10^{-2}$ & $2.77 \pm 0.11 \pm 0.01 \t 10^{-2}$\\ \hline

$B_c^-\J  \underline{D}_{1}^{-}$ & $  3.45 \pm 0.28 \pm 0.03 \t 10^{-5}$ & $1.42 \pm 0.06 \pm 0.01 \t 10^{-2}$ & $ 7.64 \pm 0.38 \pm 0.10 \t 10^{-3}$ & $ 1.49 \pm 0.05 \pm 0.00 \t 10^{-2}$ \\ \hline

$B_c^- \to  D^{*-} \chi _{c1}$ & $  4.95 \pm 0.33 \pm 0.03 \t 10^{-5}$ & $ 2.13 \pm 0.07 \pm 0.01 \t 10^{-2}$ & $1.60 \pm 0.05 \pm 0.00 \t 10^{-2}$ & $ 3.01 \pm 1.65 \pm 0.13 \t 10^{-4}$
 \\ \hline

$B_c^- \to  D^{*-}  h_{c1}$ & $  3.88 \pm 0.23 \pm 0.04 \t 10^{-5}$ & $2.19 \pm 0.07 \pm 0.00 \t 10^{-2}$ & $ 7.68 \pm 0.22 \pm 0.19 \t 10^{-3}$ & $ 4.87 \pm 0.15 \pm 0.13 \t 10^{-3}$  \\ \hline

$B_c^- \to  \rho ^- \bar{D}_1^0$ & $ 1.01 \pm 0.05 ^{+0.20}_{-0.11} \t 10^{-5}$ & $8.92 \pm 0.27 ^{+2.02}_{-1.02} \t 10^{-3}$ & $ 1.17 \pm 0.04 ^{+0.13}_{-0.08} \t 10^{-3}$ & $ 3.89 \pm 0.11 ^{+0.80}_{-0.50} \t 10^{-4}$ \\ \hline

$B_c^- \to  \rho ^- \underline{\bar{D}}_{1}^{0}$ & $  1.19 \pm 0.07 ^{+0.29}_{-0.14} \t 10^{-6}$ & $ 3.06 \pm 0.10 ^{+0.35}_{-0.18} \t 10^{-3}$ & $3.21 \pm 0.09 \pm 0.06 \t 10^{-4}$ & $1.93 \pm 0.06 ^{+0.50}_{-0.30} \t 10^{-4}$ \\ \hline

$B_c^- \to  \bar{D}^{*0} a_1^-$ & $  1.09 \pm 0.07 \pm 0.20 \t 10^{-6}$ & $2.43 \pm 0.07 \pm 0.22 \t 10^{-3}$ & $ 9.34 \pm 0.28 \pm 2.03 \t 10^{-5}$ & $ 1.60 \pm 0.05 \pm 0.15 \t 10^{-3}$  \\ \hline

$B_c^- \to  \bar{D}^{*0} b_1^-$ & $  7.38 \pm 0.44 \pm 1.40 \t 10^{-12}$ & $6.42 \pm 0.18 \pm 0.63 \t 10^{-6}$ & $ 2.47 \pm 0.07 \pm 0.49 \t 10^{-7}$ & $ 4.22 \pm 0.12 \pm 0.40 \t 10^{-6}$ \\ \hline

$B_c^- \to  \rho ^0 D_1^-$ & $  7.71 \pm 2.23 ^{+1.30}_{-0.66} \t 10^{-8}$ & $7.73 \pm 1.13 ^{+0.75}_{-0.40} \t 10^{-4}$ & $ 1.01\pm 0.15 ^{+0.10}_{-0.06} \t 10^{-4}$ & $ 4.58 \pm 0.67 ^{+0.60}_{-0.41} \t 10^{-5}$ \\ \hline

$B_c^- \to  \rho ^0 \underline{D}_{1}^{-}$ & $  8.78 \pm 2.53 ^{+1.66}_{-0.80} \t 10^{-9}$ & $ 2.65 \pm 0.39 ^{+0.30}_{-0.15} \t 10^{-4}$ & $2.79 \pm 0.41 \pm 0.05 \t 10^{-5}$ & $1.67 \pm 0.25 ^{+0.37}_{-0.26} \t 10^{-5}$ \\ \hline

$B_c^- \to  \phi  D_1^-$ & $  7.84 \pm 2.26 ^{+1.35}_{-0.70} \t 10^{-8}$ & $7.77 \pm 1.14 ^{+0.70}_{-0.40} \t 10^{-4}$ & $ 1.36 \pm 0.20 ^{+0.16}_{-0.09} \t 10^{-4}$ & $ 6.57 \pm 4.58 ^{+0.67}_{-0.56} \t 10^{-5}$
 \\ \hline

$B_c^- \to  \phi  \underline{D}_{1}^{-}$ & $  8.78 \pm 2.53 ^{+1.50}_{-0.78} \t 10^{-9}$ & $ 2.60 \pm 0.38 ^{+0.28}_{-0.14} \t 10^{-4}$ & $3.74 \pm 0.55 \pm 0.05 \t 10^{-5}$ & $2.24 \pm 0.33 ^{+0.45}_{-0.36} \t 10^{-5}$ \\ \hline

$B_c^- \to  D^{*-} a_1^0$ & $  8.16 \pm 2.34 \pm 1.14 \t 10^{-9}$ & $2.09 \pm 0.31 \pm 0.19 \t 10^{-4}$ & $8.03 \pm 1.18 \pm 1.50 \t 10^{-6}$ & $ 1.37 \pm 0.20 \pm 0.01 \t 10^{-4}$  \\ \hline

$B_c^- \to  D^{*-} f_1$ & $  8.23 \pm 2.36 \pm 1.15 \t 10^{-9}$ & $ 2.07 \pm 0.30 \pm 0.18 \t 10^{-4}$ & $8.28 \pm 1.21 \pm 1.55 \t 10^{-6}$ & $1.42 \pm 0.21 \pm 0.11 \t 10^{-4}$ \\ \hline

$B_c^- \to  \omega  D_1^-$ & $  4.86 \pm 1.40 ^{+0.80}_{-0.30} \t 10^{-9}$ & $1.94 \pm 0.29 ^{+0.20}_{-0.10} \t 10^{-4}$ & $ 2.56 \pm 0.38 ^{+0.51}_{-0.15} \t 10^{-5}$ & $8.34 \pm 1.03 ^{+1.50}_{-1.00} \t 10^{-6}$
 \\ \hline

$B_c^- \to  \omega  \underline{D}_{1}^{-}$ & $ 5.68 \pm 1.64 ^{+1.05}_{-0.50} \t 10^{-10}$ & $ 6.62 \pm 0.98 ^{+0.70}_{-0.36} \t 10^{-5}$ & $7.08 \pm 1.04 \pm 0.12 \t 10^{-7}$ & $4.23 \pm 0.62 ^{+0.95}_{-0.67} \t 10^{-6}$ \\ \hline

\end{tabular}
\end{table}

\begin{table}
\captionof{table} {Branching ratios and helicity amplitudes of $B_{c}  \to  VA$ decays for CKM-suppressed ($\Delta  b = 1, \Delta C = -1, \Delta S = 0$) mode.}
\label{t9}
\footnotesize
\begin{tabular}{c c c c c} \hline
\multirow{2}{*}{Decays} &  \multirow{2}{*}{Branching Ratios} & \multicolumn{3}{c}{Helicity Amplitudes} \\ \cline{3-5}
 \multicolumn{2}{c}{}& $|H_0|$& $|H_+|$& $|H_-|$ \\  \hline

$B_c^- \to  D^{*-} \bar{D}_1^0$ & $  8.13 \pm 0.51 ^{+0.96}_{-1.48} \t 10^{-7}$ & $ 2.57 \pm 0.08 ^{+0.26}_{-0.16} \t 10^{-3}$ & $9.39 \pm 0.29 ^{+0.60}_{-0.67} \t 10^{-4}$ & $3.98 \pm 0.18 ^{+0.50}_{-0.43} \t 10^{-4}$ \\ \hline

$B_c^- \to  D^{*-} \underline{\bar{D}}_{1}^{0}$ & $  7.01 \pm 0.48 ^{+1.51}_{-0.78} \t 10^{-8}$ & $7.49 \pm 0.26 ^{+0.81}_{-0.47} \t 10^{-4}$ & $ 2.55 \pm 0.08 \pm 0.04 \t 10^{-4}$ & $ 1.80 \pm 0.08 ^{+0.43}_{-0.27} \t 10^{-4}$ \\ \hline

$B_c^- \to  \bar{D}^{*0} D_1^-$ & $ 7.09 \pm 0.99 ^{+0.50}_{-0.33} \t 10^{-8}$ & $ 6.38 \pm 0.55 ^{+0.12}_{-0.01} \t 10^{-4}$ & $1.37 \pm 0.17 ^{+0.08}_{-0.02} \t 10^{-4}$ & $4.88 \pm 0.19 \pm 0.35 \t 10^{-4}$
 \\ \hline

$B_c^- \to  \bar{D}^{*0} \underline{D}_{1}^{-}$& $  1.57 \pm 0.17 ^{+0.02}_{-0.01} \t 10^{-7}$ & $2.77 \pm 0.23 ^{+0.11}_{-0.09} \t 10^{-4}$ & $ 4.34 \pm 0.49 ^{+0.33}_{-0.24} \t 10^{-5}$ & $ 2.67 \pm 0.10 \pm 0.19 \t 10^{-4}$  \\ \hline

\end{tabular}
\end{table}
\begin{table}
\captionof{table} {Branching ratios and helicity amplitudes of $B_{c}  \to  VA$ decays for CKM-suppressed ($\Delta  b = 1, \Delta C = -1, \Delta S = -1$) mode.}
\label{t10}
\footnotesize
\begin{tabular}{c c c c c} \hline
\multirow{2}{*}{Decays} &  \multirow{2}{*}{Branching Ratios} & \multicolumn{3}{c}{Helicity Amplitudes} \\ \cline{3-5}
 \multicolumn{2}{c}{}& $|H_0|$& $|H_+|$& $|H_-|$ \\  \hline

$B_c^- \to  D_s^{*-} \bar{D}_1^0$ & $  2.01 \pm 0.13 ^{+0.44}_{-024.} \t 10^{-5}$ & $1.28 \pm 0.04 ^{+0.14}_{-0.08} \t 10^{-2}$ & $ 4.94 \pm 0.15 ^{+0.55}_{-0.40} \t 10^{-3}$ & $ 2.25 \pm 0.12 ^{+0.33}_{-0.24} \t 10^{-3}$ \\ \hline

$B_c^- \to  D_s^{*-} \underline{\bar{D}}_{1}^{0}$ &
$ 1.71 \pm 0.12 ^{+0.37}_{-0.19} \t 10^{-6}$ & $ 3.68 \pm 0.13 ^{+0.40}_{-0.23} \t 10^{-3}$ & $1.34 \pm 0.04 \pm 0.02 \t 10^{-3}$ & $9.85 \pm 0.25 ^{+0.21}_{-0.14} \t 10^{-4}$ \\ \hline

$B_c^- \to  \bar{D}^{*0} D_{s1}^-$ & $  1.42 \pm 0.22 \pm 0.10 \t 10^{-6}$ & $2.99 \pm 0.28 \pm 0.09 \t 10^{-3}$ & $ 6.76 \pm 0.89 \pm 0.02 \t 10^{-4}$ & $ 2.00 \pm 0.08 \pm 0.15 \t 10^{-3}$
 \\ \hline

$B_c^- \to  \bar{D}^{*0} \underline{D}_{s1}^{-}$  & $  3.02 \pm 0.31 \pm 0.39 \t 10^{-7}$ & $ 1.19 \pm 0.08 \pm 0.18 \t 10^{-3}$ & $2.67 \pm 0.31 \pm 0.17 \t 10^{-4}$ & $1.19 \pm 0.02 \pm 0.04 \t 10^{-3}$  \\ \hline

\end{tabular}
\end{table}
\begin{table}
\captionof{table} {Branching ratios and helicity amplitudes of $B_{c}  \to  AA$ decays for CKM-favored ($\Delta  b = 1, \Delta C = 1, \Delta S = 0$) mode.}
\label{t11}
\footnotesize
\begin{tabular}{c c c c c} \hline
\multirow{2}{*}{Decays} &  \multirow{2}{*}{Branching Ratios} & \multicolumn{3}{c}{Helicity Amplitudes} \\ \cline{3-5}
 \multicolumn{2}{c}{}& $|H_0|$& $|H_+|$& $|H_-|$ \\  \hline

$B_c^-  \to  \chi _{c1} a_1^-$ & $ 1.81 \pm 0.11 \pm 0.01 \t 10^{-4}$ & $2.15 \pm 0.06 \pm 0.17 \t 10^{-4}$ & $6.20 \pm 0.19 \pm 0.52 \t 10^{-3}$ & $ 4.36 \pm 0.13 \pm 0.02 \t 10^{-2}$ \\ \hline

$B_c^-  \to  b_1^- \chi _{c1}$ & $ 1.26 \pm 0.07 \pm 0.01 \t 10^{-9}$ & $5.61 \pm 0.17 \pm 0.45 \t 10^{-7}$ & $1.64 \pm 0.05 \pm 0.13 \t 10^{-5}$ & $ 1.15 \pm 0.03 \pm 0.00 \t 10^{-4}$ \\ \hline

$B_c^-  \to  h_{c1} a_1^-$ & $ 1.36 \pm 0.08 \pm 0.10 \t 10^{-4}$ & $3.32 \pm 0.10 \pm 0.19 \t 10^{-2}$ & $8.57 \pm 0.25 \pm 0.29 \t 10^{-3}$ & $1.72 \pm 0.05 \pm 0.05 \t 10^{-2}$\\ \hline

$B_c^-  \to  h_{c1} b_1^-$ & $ 9.52 \pm 0.58 \pm 0.67 \t 10^{-10}$ & $8.77 \pm 0.26 \pm 0.50 \t 10^{-5}$ & $2.27 \pm 0.07 \pm 0.08 \t 10^{-5}$ & $4.55 \pm 0.14 \pm 0.13 \t 10^{-5}$\\ \hline

$B_c^-  \to  D_1^- D_1^0$ & $ 3.10 \pm 0.90 ^{+0.81}_{-0.49} \t 10^{-6}$ & $4.18 \pm 0.61 ^{+0.72}_{-0.51} \t 10^{-3}$ & $ 1.33 \pm 0.19 ^{+0.23}_{-0.16} \t 10^{-3}$ & $ 3.50 \pm 0.50 ^{+0.41}_{-0.23} \t 10^{-3}$
 \\ \hline
$B_c^-  \to  \underline{D}_{1}^{0} D_1^-$ & $ 1.00 \pm 0.30 ^{+0.25}_{-0.15} \t 10^{-6}$ & $ 2.33 \pm 0.34 ^{+0.41}_{-0.28} \t 10^{-3}$ & $7.42 \pm1.09 \pm 0.10 \t 10^{-4}$ & $1.95 \pm 0.28 ^{+0.21}_{-0.13} \t 10^{-3}$ \\ \hline

$B_c^-  \to  \underline{D}_{1}^{-} D_1^0$ & $ 3.66 \pm 1.01 ^{+0.60}_{-0.30} \t 10^{-7}$ & $ 1.60 \pm 0.23 ^{+0.14}_{-0.09} \t 10^{-3}$ & $5.53 \pm0.91 ^{+1.26}_{-0.95} \t 10^{-4}$ & $0.94 \pm 0.14 ^{+0.02}_{-0.00} \t 10^{-3}$ \\ \hline

$B_c^-  \to  \underline{D}_{1}^{-} \underline{D}_{1}^{0}$ & $ 1.14 \pm 0.33 ^{+0.18}_{-0.09} \t 10^{-7}$ & $0.89 \pm 0.13 ^{+0.08}_{-0.05} \t 10^{-3}$ & $ 3.08 \pm 0.45^{+0.71}_{-0.53} \t 10^{-4}$ & $ 5.20 \pm 0.76^{+0.10}_{-0.03} \t 10^{-4}$ \\ \hline

\end{tabular}
\end{table}
\begin{table}
\captionof{table} {Branching ratios and helicity amplitudes of $B_{c}  \to  AA$ decays for CKM-favored ($\Delta  b = 1, \Delta C = 0, \Delta S = -1$) mode.}
\label{t12}
\footnotesize
\begin{tabular}{c c c c c} \hline
\multirow{2}{*}{Decays} &  \multirow{2}{*}{Branching Ratios} & \multicolumn{3}{c}{Helicity Amplitudes} \\ \cline{3-5}
 \multicolumn{2}{c}{}& $|H_0|$& $|H_+|$& $|H_-|$ \\  \hline

$B_c^-  \to  h_{c1} D_{s1}^-$ &  $ 3.15 \pm 0.19 \pm 0.18 \t 10^{-5}$ & $ 5.59 \pm 0.17 \pm 0.79 \t 10^{-3}$ & $1.40 \pm 0.46 \pm 0.03 \t 10^{-2}$ & $1.96 \pm 0.06 \pm 0.04 \t 10^{-2}$
 \\ \hline

$B_c^-  \to  \chi _{c1} D_{s1}^-$ & $ 1.22 \pm 0.15 \pm 0.05 \t 10^{-4}$ & $1.66 \pm 0.15 \pm 0.15 \t 10^{-2}$ & $ 2.91 \pm 0.71 \pm 1.11 \t 10^{-3}$ & $ 4.51 \pm 0.27 \pm 0.05 \t 10^{-2}$
\\ \hline
$B_c^-  \to  \underline{D}_{s1}^{-} \chi _{c1}$ & $ 2.88 \pm 0.27 \pm 0.06 \t 10^{-5}$ & $ 8.28 \pm 0.54 \pm 0.58 \t 10^{-3}$ & $ 2.98 \pm 0.72 \pm 0.10 \t 10^{-3}$ & $2.31 \pm 0.11 \pm 0.01 \t 10^{-2}$
 \\ \hline

$B_c^-  \to  h_{c1} \underline{D}_{s1}^{-}$ & $ 1.17 \pm 0.07 \pm 0.06 \t 10^{-5}$ & $4.89 \pm 0.15 \pm 0.44 \t 10^{-3}$ & $ 8.93 \pm 0.27 \pm 0.21 \t 10^{-3}$ & $ 1.21 \pm 0.04 \pm 0.02 \t 10^{-2}$
\\ \hline
$B_c^-  \to  \underline{K}_{1}^{-} \bar{D}_1^0$ & $ 1.90 \pm 0.12 ^{+0.70}_{-0.42} \t 10^{-7}$ & $1.16 \pm 0.03 ^{+0.24}_{-0.15} \t 10^{-3}$ & $ 1.73 \pm 0.05 ^{+0.33}_{-0.23} \t 10^{-4}$ & $ 5.10 \pm 0.15 ^{+0.59}_{-0.33} \t 10^{-4}$\\ \hline

$B_c^-  \to  \underline{\bar{D}}_{1}^{0} K_1^-$ & $ 1.30 \pm 0.08 ^{+0.23}_{-0.10} \t 10^{-8}$ & $3.08 \pm 0.09 ^{+0.26}_{-0.14} \t 10^{-4}$ & $ 6.18 \pm 0.19 ^{+1.50}_{-1.12} \t 10^{-4}$ & $ 1.04 \pm 0.04 \pm 0.02 \t 10^{-4}$
\\ \hline

$B_c^-  \to  \bar{D}_1^0 K_1^-$ & $ 1.22 \pm 0.07 ^{+0.48}_{-0.27} \t 10^{-7}$ & $ 9.35 \pm 0.28 ^{+1.75}_{-1.24} \t 10^{-4}$ & $1.25 \pm 0.04 ^{+0.23}_{-0.17} \t 10^{-4}$ & $3.70 \pm 0.11 ^{+0.42}_{-0.24} \t 10^{-4}$\\ \hline

$B_c^-  \to  \underline{\bar{D}}_{1}^{0} \underline{K}_{1}^{-}$ & $ 2.00 \pm 0.10 ^{+0.34}_{-0.15} \t 10^{-8}$ & $ 3.78 \pm 0.12 ^{+0.32}_{-0.17} \t 10^{-4}$ & $8.39 \pm 0.25 ^{+0.20}_{-0.15} \t 10^{-5}$ & $1.41 \pm 0.05 \pm 0.03 \t 10^{-4}$\\ \hline

$B_c^-  \to  \underline{D}_{s1}^{-} a_1^0$ & $ 2.84\pm 0.82 \pm 0.09 \t 10^{-10}$ & $ 4.57 \pm 0.68 \pm 0.10 \t 10^{-5}$ & $4.51 \pm 0.66 \pm 0.75 \t 10^{-6}$ & $1.62 \pm 0.24 \pm 0.02 \t 10^{-5}$
\\ \hline
$B_c^-  \to  \underline{D}_{s1}^{-} f_1$ & $ 2.75 \pm 0.79 \pm 0.90 \t 10^{-10}$ & $4.49 \pm 0.66 \pm 0.10 \t 10^{-5}$ & $ 4.71 \pm 0.70\pm 0.75 \t 10^{-6}$ & $ 1.65 \pm 0.25 \pm 0.03 \t 10^{-5}$\\ \hline

$B_c^-  \to  D_{s1}^- a_1^0$ & $ 1.02 \pm 0.29 \pm 0.16 \t 10^{-9}$ & $7.88 \pm 1.16 \pm 0.90 \t 10^{-5}$ & $ 1.15 \pm 0.17 \pm 0.12 \t 10^{-5}$ & $ 4.54 \pm 0.67 \pm 0.16 \t 10^{-5}$
\\ \hline
$B_c^-  \to  D_{s1}^- f_1$ & $ 1.02 \pm 0.29 \pm 0.16 \t 10^{-9}$ & $ 7.83 \pm 1.16 \pm 0.90 \t 10^{-5}$ & $1.20 \pm 0.18 \pm 0.13 \t 10^{-5}$ & $4.71 \pm 0.69\pm 0.16 \t 10^{-5}$
\\ \hline

\end{tabular}
\end{table}
\begin{table}
\captionof{table} {Branching ratios and helicity amplitudes of $B_{c}  \to  AA$ decays for CKM-suppressed ($\Delta  b = 1, \Delta C = 1, \Delta S = -1$) mode.}
\label{t13}
\footnotesize
 \begin{tabular}{c c c c c}\hline
\multirow{2}{*}{Decays} &  \multirow{2}{*}{Branching Ratios} & \multicolumn{3}{c}{Helicity Amplitudes} \\ \cline{3-5}
 \multicolumn{2}{c}{}& $|H_0|$& $|H_+|$& $|H_-|$ \\  \hline

$B_c^-  \to  \underline{K}_{1}^{-} \chi _{c1}$& $ 1.18 \pm 0.07 \pm 0.01 \t 10^{-5}$ & $3.27 \pm 0.10 \pm 0.03 \t 10^{-4}$ & $1.61 \pm 0.05 \pm 0.14 \t 10^{-3}$ & $ 1.14 \pm 0.03 \pm 0.00 \t 10^{-2}$ \\ \hline
$B_c^-  \to  \chi _{c1} K_1^-$ & $ 6.76 \pm 0.40 \pm 0.04 \t 10^{-6}$ & $ 9.26 \pm 0.28 \pm 1.70 \t 10^{-5}$ & $ 1.21 \pm 0.04 \pm 0.10 \t 10^{-3}$ & $8.48 \pm 0.26 \pm 0.01 \t 10^{-3}$\\ \hline
$B_c^-  \to  h_{c1} K_1^-$ & $ 4.99 \pm 0.30 \pm 0.34 \t 10^{-6}$ & $ 6.31 \pm 0.19 \pm 0.36 \t 10^{-3}$ & $ 1.73 \pm 0.05 \pm 0.05 \t 10^{-3}$ & $ 3.42 \pm 0.10 \pm 0.10 \t 10^{-3}$ \\ \hline
$B_c^-  \to  h_{c1} \underline{K}_{1}^{-}$ & $ 6.63 \pm 0.40 \pm 0.45 \t 10^{-6}$ & $6.91 \pm 0.21 \pm 0.44 \t 10^{-3}$ & $2.36 \pm 0.07 \pm 0.08 \t 10^{-3}$ & $4.59 \pm 0.13 \pm 0.13 \t 10^{-3}$ \\ \hline
$B_c^-  \to  D_{s1}^- D_1^0$ &  $ 2.95 \pm 1.05 ^{+0.20}_{-0.16} \t 10^{-7}$ & $7.67 \pm 1.14 ^{+0.96}_{-0.84} \t 10^{-4}$ & $ 2.75 \pm 0.41 ^{+0.30}_{-0.27} \t 10^{-4}$ & $ 0.91 \pm 0.13 ^{+0.04}_{-0.04} \t 10^{-3}$ \\ \hline
$B_c^-  \to  \underline{D}_{1}^{0} D_{s1}^-$ & $ 4.57 \pm 1.32 \pm 0.60 \t 10^{-8}$ & $ 4.28 \pm 0.63 \pm 0.50 \t 10^{-4}$ & $1.52 \pm 0.22 \pm 0.16 \t 10^{-4}$ & $5.11 \pm 0.75 \pm 0.19 \t 10^{-4}$\\ \hline
$B_c^-  \to  \underline{D}_{s1}^{-} D_1^0$ & $ 2.85 \pm 0.82 \pm 0.11 \t 10^{-8}$ & $ 4.46 \pm 0.66 \pm 0.15 \t 10^{-4}$ & $7.80 \pm 1.45 \pm 1.60 \t 10^{-5}$ & $3.07 \pm 0.45 \pm 1.04 \t 10^{-4}$ \\ \hline
$B_c^-  \to  \underline{D}_{s1}^{-} \underline{D}_{1}^{0}$ & $ 8.92 \pm 2.58 \pm 0.30 \t 10^{-9}$ & $2.49 \pm 0.37 \pm 0.08 \t 10^{-4}$ & $ 4.34 \pm 0.64 \pm 0.90 \t 10^{-5}$ & $ 1.71 \pm 0.25 \pm 0.02 \t 10^{-4}$
 \\ \hline

\end{tabular}
\end{table}
\begin{table}
\captionof{table} {Branching ratios and helicity amplitudes of $B_{c}  \to  AA$ decays for CKM-suppressed ($\Delta  b = 1, \Delta C = 0, \Delta S = 0$) mode.}
\label{t14}
\footnotesize
\begin{tabular}{c c c c c}\hline
\multirow{2}{*}{Decays} &  \multirow{2}{*}{Branching Ratios} & \multicolumn{3}{c}{Helicity Amplitudes} \\ \cline{3-5}
 \multicolumn{2}{c}{}& $|H_0|$& $|H_+|$& $|H_-|$ \\  \hline

$B_c^-  \to  \underline{D}_{1}^{-} \chi _{c1}$ & $ 1.60 \pm 0.15 \pm 0.04 \t 10^{-6}$ & $1.78 \pm 0.13 \pm 0.20 \t 10^{-3}$ & $4.08 \pm 2.66 \pm 0.95 \t 10^{-5}$ & $ 5.28 \pm 0.22 \pm 0.02 \t 10^{-3}$
 \\ \hline
$B_c^-  \to  \chi _{c1} D_1^-$ & $ 7.48 \pm 0.92 \pm 0.50 \t 10^{-6}$ & $ 4.43 \pm 0.66 \pm 0.46 \t 10^{-3}$ & $5.87 \pm 0.67 \pm 0.36 \t 10^{-4}$ & $1.52 \pm 0.49 \pm 0.02 \t 10^{-2}$
\\ \hline
$B_c^-  \to  h_{c1} D_1^-$ & $ 1.79 \pm 0.11 \pm 0.01 \t 10^{-6}$ & $1.12 \pm 0.03 \pm 0.20 \t 10^{-3}$ & $ 3.26 \pm 0.09 \pm 0.08 \t 10^{-3}$ & $ 4.61 \pm 0.10 \pm 0.11 \t 10^{-3}$
\\ \hline
$B_c^-  \to  h_{c1} \underline{D}_{1}^{-}$ & $ 5.59 \pm 0.33 \pm 0.32 \t 10^{-7}$ & $ 6.04 \pm 0.19 \pm 1.10 \t 10^{-4}$ & $1.81 \pm 0.05 \pm 0.04 \t 10^{-3}$ & $2.60 \pm 0.08 \pm 0.06 \t 10^{-3}$ \\ \hline
$B_c^-  \to  \bar{D}_1^0 a_1^-$ & $ 3.41 \pm 0.20 ^{+1.30}_{-0.77}. \t 10^{-6}$ & $4.96 \pm 0.14 ^{+0.94}_{-0.66} \t 10^{-3}$ & $ 6.35 \pm 0.20 ^{+1.22}_{-0.84} \t 10^{-4}$ & $ 1.89 \pm 0.06 ^{+0.22}_{-0.12} \t 10^{-3}$
\\ \hline
$B_c^-  \to  b_1^- \bar{D}_1^0$ & $ 2.38 \pm 0.14 ^{+0.92}_{-0.54} \t 10^{-11}$ & $1.35 \pm 0.04 ^{+0.25}_{-0.17} \t 10^{-5}$ & $ 1.68 \pm 0.05 ^{+0.32}_{-0.22} \t 10^{-6}$ & $ 4.95 \pm 0.10 ^{+0.58}_{-0.33} \t 10^{-6}$
\\ \hline
$B_c^-  \to  \underline{\bar{D}}_{1}^{0} a_1^-$ & $ 3.38 \pm 0.21 ^{+0.58}_{-0.26} \t 10^{-7}$ & $ 1.57 \pm 0.05 ^{+0.13}_{-0.06} \t 10^{-3}$ & $3.12 \pm 0.09 ^{+0.76}_{-0.56} \t 10^{-4}$ & $5.22 \pm 0.15 ^{+0.10}_{-0.08} \t 10^{-4}$
\\ \hline
$B_c^-  \to  \underline{\bar{D}}_{1}^{0} b_1^-$ & $ 2.47 \pm 0.04 ^{+0.41}_{-0.18} \t 10^{-12}$ & $ 4.15 \pm 0.13 ^{+0.36}_{-0.18} \t 10^{-6}$ & $8.25 \pm 0.25 ^{+1.95}_{-1.46} \t 10^{-7}$ & $1.38 \pm 0.04 \pm 0.03 \t 10^{-6}$
 \\ \hline
$B_c^-  \to  D_1^- a_1^0$ & $ 2.61 \pm 0.75 ^{+0.80}_{-0.45} \t 10^{-8}$ & $4.29 \pm 0.64 ^{+0.75}_{-0.51} \t 10^{-4}$ & $ 5.51 \pm 0.81 ^{+0.95}_{-0.66} \t 10^{-5}$ & $ 1.62 \pm 0.22 ^{+0.17}_{-0.10} \t 10^{-4}$ \\ \hline
$B_c^-  \to  D_1^- f_1$ & $ 2.60 \pm 0.75 ^{+0.80}_{-0.45} \t 10^{-8}$ & $4.27 \pm 0.63 ^{+0.75}_{-0.51} \t 10^{-4}$ & $5.74 \pm 0.84 ^{+0.95}_{-0.66} \t 10^{-5}$ & $1.70 \pm 0.25 ^{+0.17}_{-0.10} \t 10^{-4}$
\\ \hline
$B_c^-  \to  \underline{D}_{1}^{-} a_1^0$ & $ 2.60 \pm 0.75 ^{+0.37}_{-0.16} \t 10^{-9}$ & $ 1.36 \pm 0.20 ^{+0.10}_{-0.05} \t 10^{-4}$ & $2.70 \pm 0.39 ^{+0.60}_{-0.43} \t 10^{-5}$ & $4.53 \pm 0.67 \pm 0.08 \t 10^{-5}$ \\ \hline
$B_c^-  \to  \underline{D}_{1}^{-} f_1$ & $ 2.50 \pm 0.72 ^{+0.37}_{-0.16} \t 10^{-9}$ & $1.33 \pm 0.19 ^{+0.10}_{-0.05} \t 10^{-4}$ & $ 2.79 \pm 0.41 ^{+0.60}_{-0.43} \t 10^{-5}$ & $ 4.63 \pm 0.68 \pm 0.08 \t 10^{-5}$\\ \hline

\end{tabular}
\end{table}

\begin{table}
\captionof{table} {Branching ratios and helicity amplitudes of $B_{c}  \to  AA$ decays for CKM-suppressed ($\Delta  b = 1, \Delta C = -1, \Delta S = 0$) mode.}
\label{t15}
\footnotesize
  \begin{tabular}{c c c c c} \hline
\multirow{2}{*}{Decays} &  \multirow{2}{*}{Branching Ratios} & \multicolumn{3}{c}{Helicity Amplitudes} \\ \cline{3-5}
 \multicolumn{2}{c}{}& $|H_0|$& $|H_+|$& $|H_-|$ \\  \hline
$B_c^-  \to  \bar{D}_1^0 D_1^-$ & $ 1.00 \pm 0.30 ^{+0.35}_{-0.21} \t 10^{-7}$ & $ 7.80 \pm 0.33 ^{+1.46}_{-1.06} \t 10^{-4}$ & $2.48 \pm 0.10 ^{+0.55}_{-0.33} \t 10^{-4}$ & $6.51 \pm 0.28 ^{+0.74}_{-0.46} \t 10^{-4}$\\ \hline
$B_c^-  \to  \underline{D}_{1}^{-} \bar{D}_1^0$& $ 3.00 \pm 0.23 ^{+1.00}_{-0.60} \t 10^{-8}$ & $4.20 \pm 0.16 ^{+0.76}_{-0.55} \t 10^{-4}$ & $ 1.34 \pm 0.05 ^{+0.26}_{-0.19} \t 10^{-4}$ & $ 3.41 \pm 0.12 ^{+0.38}_{-0.23} \t 10^{-4}$\\ \hline
$B_c^-  \to  \underline{\bar{D}}_{1}^{0} D_1^-$ & $ 1.42 \pm 0.14 ^{+0.29}_{-0.16} \t 10^{-8}$ & $2.12 \pm 0.15 ^{+0.34}_{-0.22} \t 10^{-4}$ & $ 1.08 \pm 0.06 ^{+0.26}_{-0.20} \t 10^{-4}$ & $ 1.95 \pm 0.10 ^{+0.08}_{-0.01} \t 10^{-4}$
\\ \hline
$B_c^-  \to  \underline{\bar{D}}_{1}^{0} \underline{D}_{1}^{-}$ & $ 3.91 \pm 0.35 ^{+0.74}_{-0.38} \t 10^{-9}$ & $ 1.66 \pm 0.07 ^{+0.16}_{-0.10} \t 10^{-4}$ & $5.75 \pm 0.25 ^{+1.46}_{-1.10} \t 10^{-5}$ & $9.69 \pm 0.41 ^{+0.20}_{-0.10} \t 10^{-5}$\\ \hline
\end{tabular}
\end{table}
\begin{table}
\captionof{table} {Branching ratios and helicity amplitudes of $B_{c}  \to  AA$ decays for CKM-suppressed ($\Delta  b = 1, \Delta C = -1, \Delta S = -1$) mode.}
\label{t16}
\footnotesize
\begin{tabular}{c c c c c}\hline
\multirow{2}{*}{Decays} &  \multirow{2}{*}{Branching Ratios} & \multicolumn{3}{c}{Helicity Amplitudes} \\ \cline{3-5}
 \multicolumn{2}{c}{}& $|H_0|$& $|H_+|$& $|H_-|$ \\  \hline
$B_c^-  \to  D_{s1}^- \bar{D}_1^0$ & $ 1.76 \pm 0.15 ^{+0.55}_{-0.34} \t 10^{-6}$ & $3.13 \pm 0.13 ^{+0.57}_{-0.42} \t 10^{-3}$ & $ 9.96 \pm 0.12 ^{+1.85}_{-1.36} \t 10^{-4}$ & $ 2.75 \pm 0.12 ^{+0.28}_{-0.19} \t 10^{-3}$ \\ \hline
$B_c^-  \to  \underline{\bar{D}}_{1}^{0} D_{s1}^-$ & $ 2.47 \pm 0.24 ^{+0.45}_{-0.26} \t 10^{-7}$ & $ 1.28 \pm 0.05 ^{+0.13}_{-0.09} \t 10^{-3}$ & $4.43 \pm 0.20 ^{+1.00}_{-0.83} \t 10^{-4}$ & $8.55 \pm 0.50 ^{+0.21}_{-0.01} \t 10^{-4}$
\\ \hline
$B_c^-  \to  \underline{D}_{s1}^{-} \bar{D}_1^0$ & $ 6.39\pm 0.50 ^{+2.00}_{-1.20} \t 10^{-7}$ & $ 1.92 \pm 0.08 ^{+0.33}_{-0.25} \t 10^{-3}$ & $6.29 \pm 0.22 ^{+1.19}_{-0.87} \t 10^{-4}$ & $1.65 \pm 0.06 ^{+0.18}_{-0.11} \t 10^{-3}$
\\ \hline
$B_c^-  \to  \underline{D}_{s1}^{-} \underline{\bar{D}}_{1}^{0}$ & $ 8.28 \pm 0.74 ^{+1.46}_{-0.77} \t 10^{-8}$ & $7.70 \pm 0.35 ^{+0.70}_{-0.45} \t 10^{-4}$ & $ 2.59 \pm 0.10 ^{+0.65}_{-0.51} \t 10^{-4}$ & $ 4.75 \pm 0.22 ^{+0.08}_{-0.04} \t 10^{-4}$
\\ \hline
\end{tabular}
\end{table}
\begin{table}
\captionof{table} {Comparison of branching ratios with available theoretical works.}
\label{t17}
\footnotesize
\begin{tabular}{c c c c c c}\hline
Decays & This Work & \cite{25} & \cite{23} & \cite{13}& \cite{15} \\ \hline
$B_c^-  \to  \chi_{c1} \rho^-$ &$1.47 \pm 0.03 \pm 0.01 \t 10^{-3}$ &$2.19  \t 10^{-4}$ & $1.4  \t 10^{-4}$ & $2.86  \t 10^{-4}$ & $9.64  \t 10^{-5}$\\ \hline
$B_c^-  \to  h_{c1} \rho^-$ & $1.24 \pm 0.08 \pm 0.01 \t 10^{-3}$&$2.19  \t 10^{-3}$ & $9.73  \t 10^{-4}$ & $2.44  \t 10^{-3}$ & $1.24  \t 10^{-3}$\\ \hline
$B_c^-  \to  \chi_{c1} K^{*-}$ &$7.07 \pm 0.43 \pm 0.04 \t 10^{-5}$&$1.58  \t 10^{-5}$ & $8.77  \t 10^{-6}$ & $1.75  \t 10^{-5}$ & $7.01  \t 10^{-6}$\\ \hline
$B_c^-  \to  h_{c1} K^{*-}$ & $6.18 \pm 0.37 \pm 0.06 \t 10^{-5}$ &$1.23  \t 10^{-4}$ & $6.75  \t 10^{-5}$ & $1.28  \t 10^{-4}$ & $6.84  \t 10^{-5}$\\ \hline
\end{tabular}
\end{table}
\end{document}